\documentclass[preprint,12pt]{elsarticle}

\usepackage{hyperref}
\usepackage{multirow}
\usepackage{subcaption}
\usepackage{amsmath}
\usepackage{bm}
\usepackage{graphicx}

\usepackage{array}
\newcolumntype{L}[1]{>{\raggedright\let\newline\\\arraybackslash\hspace{0pt}}m{#1}}
\newcolumntype{C}[1]{>{\centering\let\newline\\\arraybackslash\hspace{0pt}}m{#1}}
\newcolumntype{R}[1]{>{\raggedleft\let\newline\\\arraybackslash\hspace{0pt}}m{#1}}

\newcommand\Top{\rule{0pt}{3ex}}
\newcommand\Bot{\rule[-1.5ex]{0pt}{0pt}}

\newcommand\wavfourth{$\mathcal{W}^4$}
\newcommand\wavfourthlift{$\mathcal{W}^4_\text{li}$}
\newcommand\wavthirdavg{$\mathcal{W}^3_\text{ai}$}

\newcounter{bla}

\makeatletter
\def\ps@pprintTitle{%
 \let\@oddhead\@empty
 \let\@evenhead\@empty
 \def\@oddfoot{}%
 \let\@evenfoot\@oddfoot}
\makeatother

\begin{document}

\begin{frontmatter}

  \title{A Parallel Data Compression Framework \\ for Large Scale 3D Scientific Data}

  \author[a]{Panagiotis Hadjidoukas\corref{author}}
  \author[a]{Fabian Wermelinger}

  \cortext[author] {Corresponding author. Current address: IBM Research, Z\"{u}rich, Switzerland. \textit{E-mail address:} hat@zurich.ibm.com}
  \address[a]{Computational Science and Engineering Laboratory, ETH Z\"{u}rich, Switzerland}

  \begin{abstract}
Large scale simulations of complex systems ranging from climate and astrophysics to crowd dynamics, produce routinely petabytes of data and are projected to reach the zettabytes level in the coming decade. These simulations enable unprecedented insights  but at the same their effectiveness is hindered by the enormous data sizes associated with the computational elements and respective output quantities of interest that impose severe constraints on storage and I/O time.
In this work, we address these challenges through a novel software framework for scientific data compression. The software (CubismZ) incorporates efficient wavelet based techniques and the state-of-the-art ZFP, SZ and FPZIP floating point compressors. The framework relies on a block-structured data layout, benefits from OpenMP and MPI and targets supercomputers based on multicores.
CubismZ can be used as a tool for ex situ (offline) compression of scientific datasets and supports conventional Computational Fluid Dynamics (CFD) file formats. Moreover, it provides a testbed of comparison, in terms of compression factor and peak signal-to-noise ratio, for a number of available data compression methods. The software yields in situ compression ratios of 100x or higher for fluid dynamics data produced by petascale simulations of cloud cavitation collapse using $\mathcal{O}(10^{11})$ grid cells, with negligible impact on the total simulation time.

  \end{abstract}


\end{frontmatter}



{\bf PROGRAM SUMMARY}

\begin{small}
\noindent
{\em Program Title: CubismZ}                                          \\
{\em Licensing provisions(please choose one): BSD 3-Clause}           \\
{\em Programming language: C++}                                   \\
%
%
{\em Nature of problem:}\\  
Large scale simulations of complex systems ranging from climate and astrophysics to crowd dynamics, produce routinely petabytes of data and are projected to reach the zettabytes level in the coming decade. These simulations enable unprecedented insights but at the same their effectiveness is hindered by the enormous data sizes associated with the computational elements and respective output quantities of interest that impose severe constraints on storage and I/O time. \\
{\em Solution method:}\\ 
We introduce CubismZ, a novel software framework and tool for data compression of 3D scientific datasets.
The software uses efficient wavelet based compression techniques and integrates the state-of-the-art ZFP [1], SZ [2] and FPZIP [3] floating point compressors. The framework relies on a block-structured data layout, benefits from OpenMP and MPI and targets supercomputers based on multicores.
CubismZ can be used as a tool for ex situ compression of scientific datasets and supports conventional Computational Fluid Dynamics file formats (HDF5 [4]).  \\
{\em Restrictions and Unusual features:}\\  
The software requires a C++ compiler, MPI and the parallel HDF5 library. \\
OpenMP, which is enabled by default, can be used for multicore compression.   \\
The input and output files must be stored in a common/shared filesystem. \\
MPI rank must be assigned equal-sized partitions of the dataset. \\
The compression scheme, the block size and the precision (float, double) are defined as compile-time options. \\
The size of the cubic data blocks must be a power of 2. \\

\end{small}

\section{Introduction}

  The continuous growth of supercomputing power allows for simulations at unprecedented resolutions, enabling unique physical insight for complex systems over on a multitude of spatio-temporal scales. In recent years  we have observed such state-of-the-art advances in many scientific areas including  cosmology~\cite{Habib:2013}, computational fluid dynamics (CFD)~\cite{Sillero:2011,Lee:2013,Rossinelli:2013}, solid earth dynamics~\cite{Rudi:2015} and atmospheric modeling~\cite{Yang:2016}.
Such large scale simulations generate massive scientific datasets that must be stored for further post processing, analysis and visualization thus imposing severe constrains on storage and I/O time.

 Today a number of high performance I/O libraries such as ADIOS~\cite{Lofstead:2008} and PnetCDF~\cite{Li:2003} have been developed for handling such large scale datasets. A recent work study~\cite{Bicer:2014} showed that data compression applied to PnetCDF can improve simulation time by up to 22\%.
  ISOBAR~\cite{Schendel:2012} has been integrated with ADIOS and, based on the FPC floating point compressor~\cite{Burtscher:2009}, was shown to achieve compression ratios between 1.9x and 52.5x for scientific datasets of less than 1 GB, on up to 2048 nodes of the Cray XK6 Jaguar cluster.
However these libraries do not include native support of data compression, while HDF5~\cite{Folk:2011} allows for generic compression but only in serial execution.
  Cubism-MPCF ~\cite{Rossinelli:2013} introduced an efficient wavelet based compression scheme combined with a comprehensive support for I/O, able to deliver up to 100x compression ratios with minimal impact on the total simulation time.
  A similar parallel data compression method for flow simulations was presented in~\cite{Sakai:2013}. The method is based on the discrete wavelet transform, followed by quantization and encoding, leading to compression ratios of up to 44x.
  The authors did not study the behavior of compression error and ratio and their largest dataset had no more than $\mathcal{O}(10^8)$ mesh points.

  GLEAN~\cite{Bui:2014} is a I/O acceleration framework for IBM BlueGene/Q supercomputers, integrated into the HACC cosmology code~\cite{Habib:2013}.
  Using topology-aware data movement, staging and compression with the ZLIB~\cite{Gailly:2004} and BLOSC~\cite{Alted:2010} lossless data compression libraries, GLEAN has shown to improve the I/O performance on BG/Q systems up to 3 times.

  ISABELA~\cite{Lakshminarasimhan:2013} is an in situ method designed to compress spatiotemporal scientific data and perform analytical operations over the compressed data.
  The method first partitions the data into small windows and applies a sorting pre-conditioner, which significantly improves the efficacy of cubic B-spline spatial compression.

  FPZIP~\cite{Lindstrom:2006} is an open-source C++ library developed at the Lawrence Livermore National Lab (LLNL) for compression of 2D or 3D datasets of floating-point scalars.
  Although designed for lossless compression, it supports lossy compression by allowing the user to specify the bits of precision.
  FPZIP is often outperformed by its successor, the ZFP compressor~\cite{Lindstrom:2014}. ZFP uses lossy but optionally error-bounded
  compression that is usually accurate to within machine epsilon in near-lossless mode, and works best for 2D and 3D arrays that exhibit spatial coherence, such as smooth fields from physics simulations and images.

  SZ~\cite{Di:2016}, developed at the Argonne National Lab (ANL), supports error-bounded data compression for large-scale HPC data sets.
  SZ outperforms other state-of-the-art compression methods and delivers compression ratios that are approximately 2x higher than those of ZFP, the second-best solution.

  In this paper, we introduce an open-source framework for parallel compression of large-scale scientific datasets, focusing on three-dimensional (3D) data produced by CFD simulations.
  The framework, named CubismZ\footnote{CubismZ is available at \url{https://github.com/phadjido/CubismZ}}.
  follows the design consideration of a hybrid MPI+OpenMP block structured flow solver that achieves unprecedented peak performance on petascale supercomputers, simulating cloud cavitation collapse of tens of thousands of bubbles~\cite{Rossinelli:2013}.
  CubismZ employs a modular scheme that utilizes a wavelet-based data compression technique and several other state-of-the-art floating point compression libraries. The main contributions of this work are the following:
  \begin{itemize}
    \item Development of CubismZ, a software framework for parallel and distributed ex situ and in situ compression of large scale 3D scientific datasets.
    \item Assessment of state-of-the-art wavelet-based data compression techniques in terms of compression ratio and distortion. 
    \item Integration of the most recent state-of-the-art floating point compressors into a common framework. 
    \item A performance study of the available compression methods for 3D cloud cavitation simulation data. 
    \item Measurements providing insight to the benefits of data compression on I/O throughput and disk storage requirements for large-scale CFD simulations running on 262'144 cores. 
  \end{itemize}

  The  paper is organized as follows: Section~\ref{sec:online} presents the software design of the proposed data compression framework. Detailed evaluation results are presented in Sections~\ref{sec:results} and~\ref{sec:performance}.  We conclude in Section~\ref{sec:conclusions}.

\section{Ex situ and in situ data compression}
  \label{sec:online}

  \subsection{Software design}
We introduce CubismZ, a standalone tool that allows for parallel compression of simulation data.
Similarly to the Cubism-MPCF flow solver~\cite{Rossinelli:2013,Hadjidoukas:2015a}, it uses a block structured grid which is initialized by loading, in parallel, the input dataset into the blocks.
  When coupled with simulation software, such as Cubism-MPCF, CubismZ serves as a module for in situ data compression.

  CubismZ can compress 3D scientific data stored in binary and HDF5 format~\cite{Folk:2011} as well as output files generated by simulation codes such as NEK5000~\cite{Fischer:2008} and NGA~\cite{Desjardins:2008}.
  The compressed files can be decompressed in parallel by our data analysis tools or even recompressed using any of the supported compression methods.
  Moreover, they can be converted to HDF5 format and visualized by tools such as Paraview or directly processed by an in-house GPU-based visualization tool.
  Following the design of Cubism-MPCF,
  CubismZ is built on top of the Cubism library, an open-source C++ framework to solve differential operators based on stencil schemes and structured uniform grids.
  The software is parallelized for multicore clusters using MPI and OpenMP and has been successfully deployed on various parallel and massively parallel architectures, including IBM BG/Q and Cray XC30, XC40 and XC50.
  CubismZ is conceptually decomposed into a cluster, node and core layer.

  The \emph{cluster layer} provides the implementation of user specific test cases (e.g.\ data readers), coordinates the progress of the data compression and implements I/O operations.
  Domain decomposition and inter-rank information exchange are based on MPI and implemented on this layer.
  The computational domain is decomposed into subdomains using a Cartesian topology with constant subdomain size.
  Each subdomain is composed of cubic grid blocks of constant size.
  The solution variables are stored on a per-cell basis which results in an Array of Structures layout.
  The constant sized grid blocks are designed to fit into the data cache of the targeted platform.
  The cluster layer is also responsible for the message exchange for the ghost cells, required by the stencil schemes used in simulations.
  This feature, however, is not required by the currently employed data compression schemes.

  The \emph{node layer} is responsible for taking advantage of the thread-level parallelism.
  Each thread processes and compresses a subset of grid blocks.
  The assignment of blocks to threads is performed using the OpenMP static scheduling policy
  with chunk size large enough to include many neighboring blocks.
  A thread processes one block at a time by copying the block data into a per-thread dedicated memory buffer.

  The \emph{core layer} provides the required data management operations, implements the 3D wavelet transforms and is interfacing with third-party lossless and lossy data compression libraries.
  Operations on the data are performed by kernels implemented on this layer and executed by a calling thread from the node layer.

  \subsection{Data compression flow}

\begin{figure*}[t]
    \begin{center}
      \includegraphics[width=0.8\textwidth]{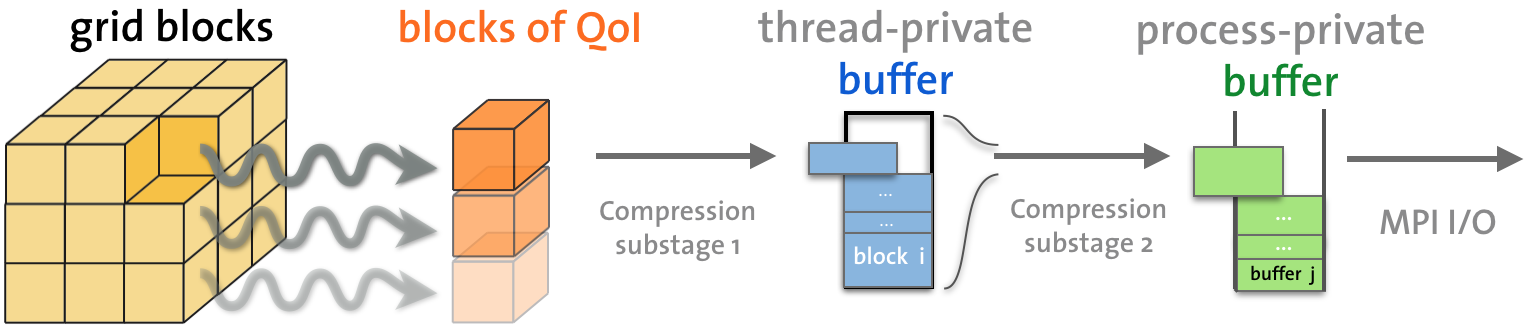}
    \end{center}
    \caption{
    Data flow of the compression scheme.
    \label{fig:waveletcompression}}
  \end{figure*}

  Figure \ref{fig:waveletcompression} provides an overview of the typical data flow in CubismZ.
  Data compression is performed at only one quantity at a time and the parallel granularity corresponds to one grid block.
  By processing one quantity at a time and performing all the compression substages in-place, whenever possible, we aim to minimize the memory requirements of the compression scheme and maximize the memory available for the simulation.

  According to the data flow, each OpenMP thread of the node layer processes a single block at a time, extracting first the field of interest (e.g.\ pressure).
  It then proceeds to the first substage of compression and writes the compressed data in a private buffer of predefined size (typically 4MB).
  When its buffer becomes full, the thread compresses the buffer and appends the output to the global buffer of the MPI process.
  The software can optionally bypass any or even both of the compression substages, applying direct data copying.

  Parallel MPI I/O is performed at the cluster layer. An exclusive prefix sum scan is performed for the determination of the file offset in the case of compression.
  Each rank acquires a destination offset and, starting from that offset, writes its compressed buffer in the file using non-collective blocking I/O write.
  Thus, our output routines generate a single file per quantity for post-processing and visualizations.

  \subsection{Main features}

  \paragraph{Wavelet-based compression} The first substage employs wavelets \cite{Donoho:1992} as they provide a balanced trade-off between compression ratio and computational cost.
  Wavelets enable data de-correlation while the separability of their associated filters leads to a very efficient forward wavelet transform.
  In terms of accuracy, it is guaranteed that the decimation will not lead to errors larger than a threshold $\varepsilon$.
  The ``on the interval'' property \cite{Cohen:1993} allows us to consider individual blocks as independent datasets that can be processed in parallel, without requiring data from neighboring blocks.

  The output of the wavelet transform is a stream of significant detail coefficients, expressed as floating point numbers, and a bit-set mask.
  Instead of encoding the results of the transform for each block independently, we concatenate them into small, per-thread buffers, as described above.
  Then, we encode each buffer as a single stream, using an external lossless coder (substage 2) such as ZLIB.
  The detail coefficients of adjacent blocks are expected to assume similar ranges, leading to more efficient data compression.

  \paragraph{Wavelet types}
  Aiming at higher compression ratios and reduced relative errors than those yielded by the initially used fourth order interpolating wavelets, we added two new types of wavelets on the interval, namely fourth order lifted and third order average interpolating wavelets.
  We also added support for data shuffling of the wavelet detail coefficients and bit zeroing of their least significant bits, enhancing subsequent lossless compression~\cite{Masui:2015}.
  Finally, we have experimented with lossless floating point compression of the detail coefficients, using the FPZIP, SZ and SPDP~\cite{SPDP:2017} compressors.
  This approach, however, does not yield better compression ratios than plain data shuffling followed by lossless compression.

  \paragraph{State-of-the-art floating point compressors}
  Besides wavelets, the first substage has been extended to include state-of-the-art floating point compressors such as ZFP, SZ, and FPZIP.
  Furthermore, we use the lossless compression of FPZIP for restart of simulations from a single compressed file containing all solution fields.

  \paragraph{Lossless compression}
  We have integrated and tested a number of lossless compression libraries in addition to ZLIB.
  For our datasets, LZMA~\cite{LZMA:2017} provides slightly better compression than ZLIB (at its highest level) but it is considerably slower.
  LZ4~\cite{LZ4:2017} offers higher compression and decompression speeds at the cost of lower compression ratios.
  ZLIB offers a good compromise between speed and compression ratio at its default compression level and, moreover, allows for in-place compression.
  Zstandard (ZSTD)~\cite{ZSTD:2017} provides compression ratio and speed comparable to that of ZLIB and LZ4, respectively.

  Lossless compression of the intermediate buffer is complement with data shuffling, which can be optionally applied to the aggregate output of the first substage.
  Thus, we shuffle both the detail coefficients and the associate bit-set masks in the case of wavelet-based compression.
  BLOSC~\cite{Alted:2010} is a high-performance lossless compression library that supports a variety of compressors, such as ZLIB and ZSTD,  along with bit and byte shuffling.
  This meta-compressor library provides an abstraction layer that simplifies the design of our framework.
  The only drawback of BLOSC is its lack of support for in-place compression.

  \paragraph{Data decompression}
   Parallel decompression is crucial for postprocessing, analysis and visualization of high-volume compressed simulation data.
   Decompression is performed at block level and applies the workflow of Cubism in the reverse order:
   first, the header (metadata) of the compressed file is accessed and the chunk that contains the target block is fetched and decompressed, according to the method used in the second substage of the compression dataflow (e.g.\ inflate() for ZLIB).
   The compressed target block stored in the chunk is then passed to the decompressor of the first substage (e.g.\ inverse wavelet transform).
   We keep recently decompressed chunks of blocks in a cache and fetch neighbor target blocks, which are also stored there due to the chunk-based processing of blocks in the first compression substage, directly from there.
   This minimizes disk accesses and avoids redundant decompressions of the same chunk.


\section{Data compression assessment}
  \label{sec:results}

  We use CubismZ to evaluate the quality of our wavelet-based compression scheme and several state-of-the-art floating point compressors on various single-precision datasets generated by our cloud cavitation simulations.
  Furthermore, we study the effect of block size to the quality and effectiveness of the data compression, as this also affects the performance of the HPC simulation software.

  Quality assessment is based on compression ratio (CR) and peak signal-to-noise-ratio (PSNR)~\cite{Lindstrom:2014,Iverson:2012}.
  CR is equal to the ratio of the uncompressed raw data size and the output compressed file size (including metadata).
  The value of PSNR for two datasets $R$ and $D$ of equal size $N$ is given by:
  \begin{equation}
    \label{eq:psnr}
    \small
    \text{PSNR} = 20 \log_{10}\left(\frac{\max_R - \min_R}{2 \sqrt{\text{$MSE_{R,D}$}}}\right)
  \end{equation}
  where $\max_R - \min_R$ is the range of values of the reference dataset $R$ and $MSE_{R,D}$ denotes the mean square error between the two datasets.
  $R$ corresponds to the original uncompressed dataset and $D$ to the dataset restored from the file generated by the lossy compression scheme for a user-specified error tolerance.
  Due to the error normalization, PSNR is a measure that allows for comparison of errors between datasets with different range of values~\cite{Iverson:2012}.
  In this paper, however, we use it to compare the performance of different compression methods for the same datasets.

  \subsection{Cloud cavitation simulation datasets}

  We compress HDF5 datasets generated by Cubism-MPCF for a simulated cloud of 70 bubbles.
  The bubbles are uniformly distributed within a sphere in a cubic domain of size $512^3$ cells, where bubble radii are drawn from a log-normal distribution.
  Each of the grid blocks contains $32^3$ points represented as scalar floating point numbers in single precision.

\begin{figure*}[!t]
    \centering
    \begin{subfigure}[t]{0.30\textwidth}
      \centering
      \resizebox{\textwidth}{!}{\includegraphics{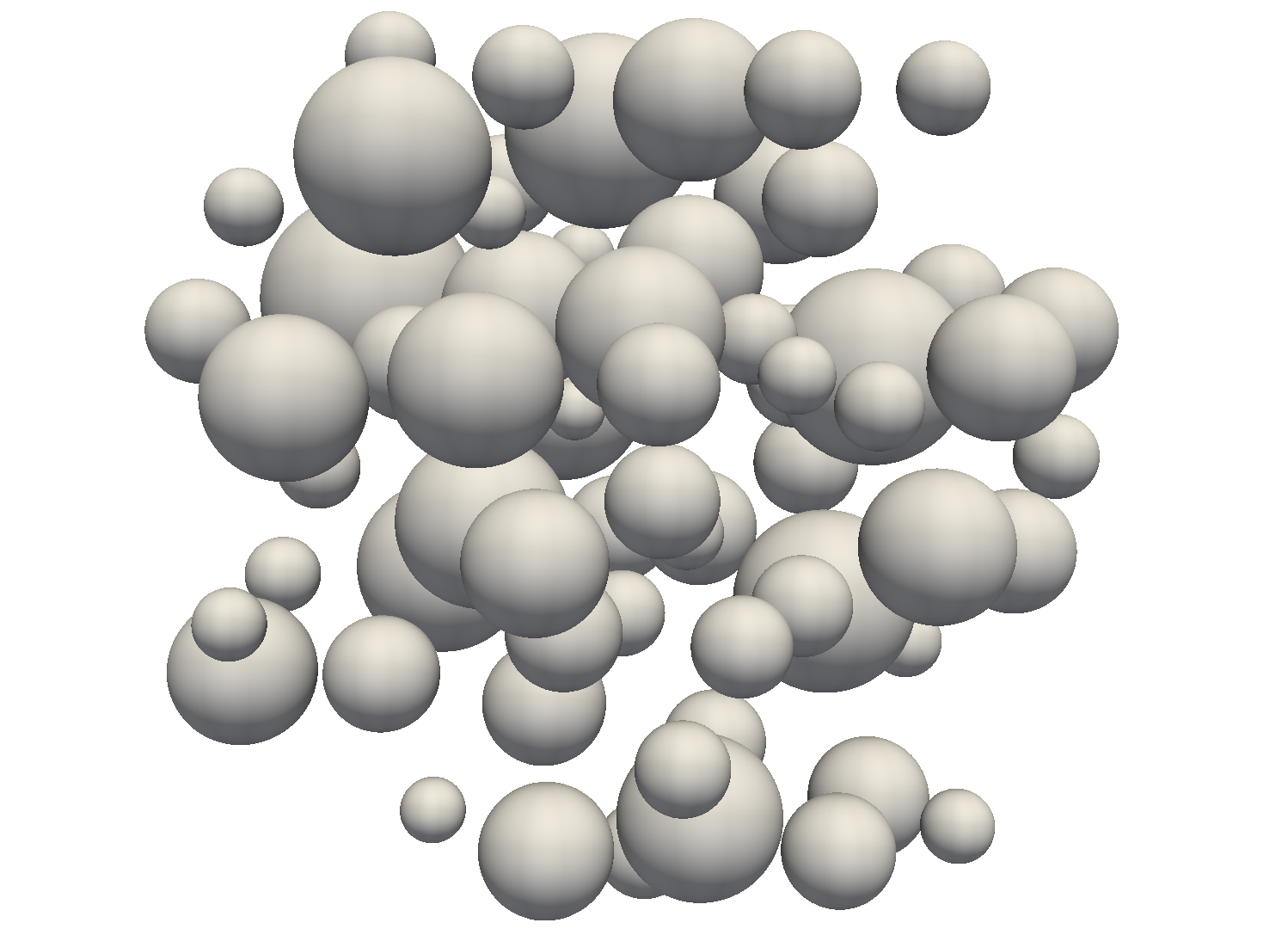}}
      \caption{Initial}
      \label{fig:sim_initial}
    \end{subfigure}%
    \hfill
    \begin{subfigure}[t]{0.30\textwidth}
      \centering
      \resizebox{\textwidth}{!}{\includegraphics{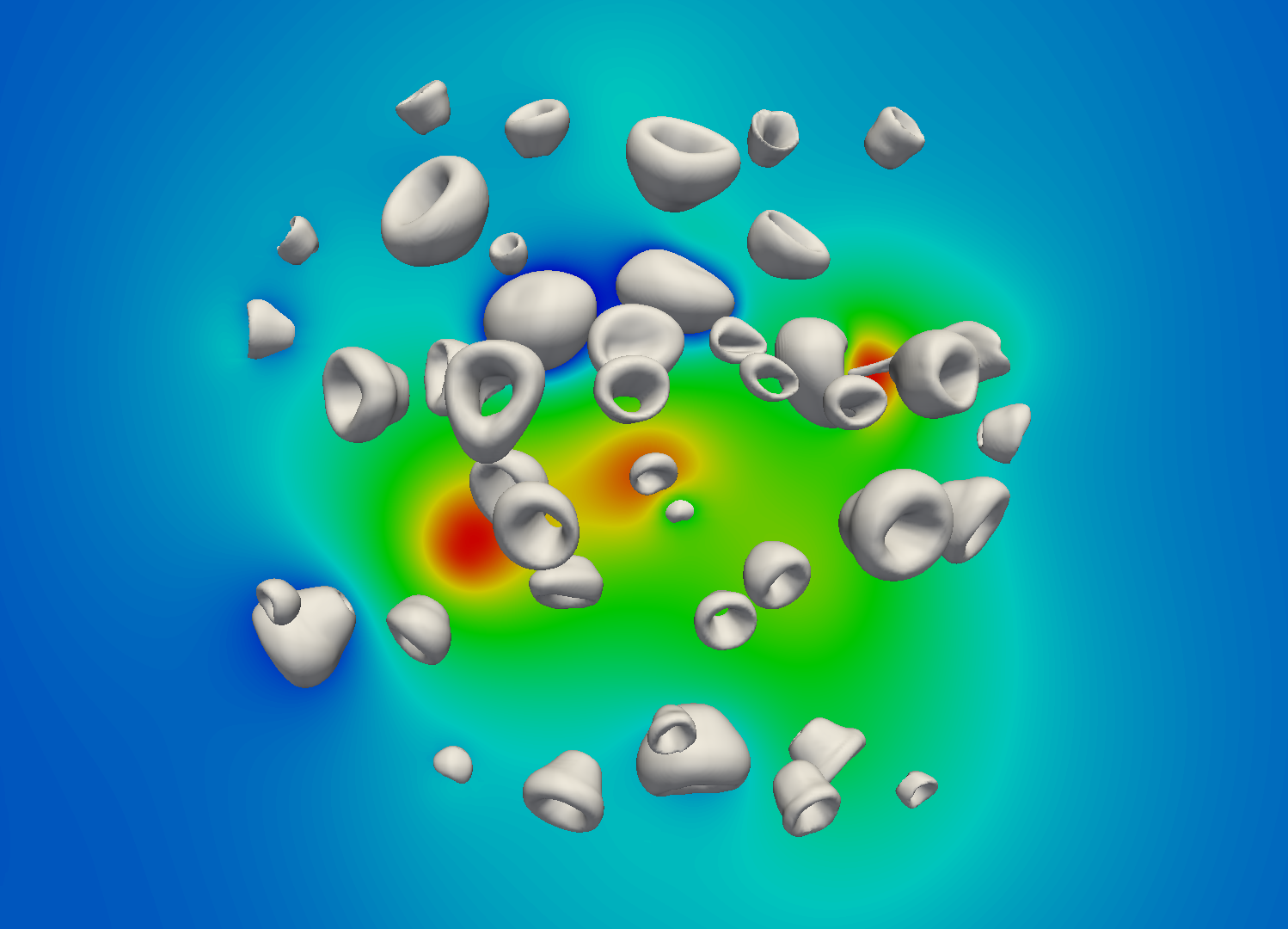}}
      \caption{$5$k steps}
      \label{fig:sim_5k_steps}
    \end{subfigure}
    \hfill
    \begin{subfigure}[t]{0.30\textwidth}
      \centering
      \resizebox{\textwidth}{!}{\includegraphics{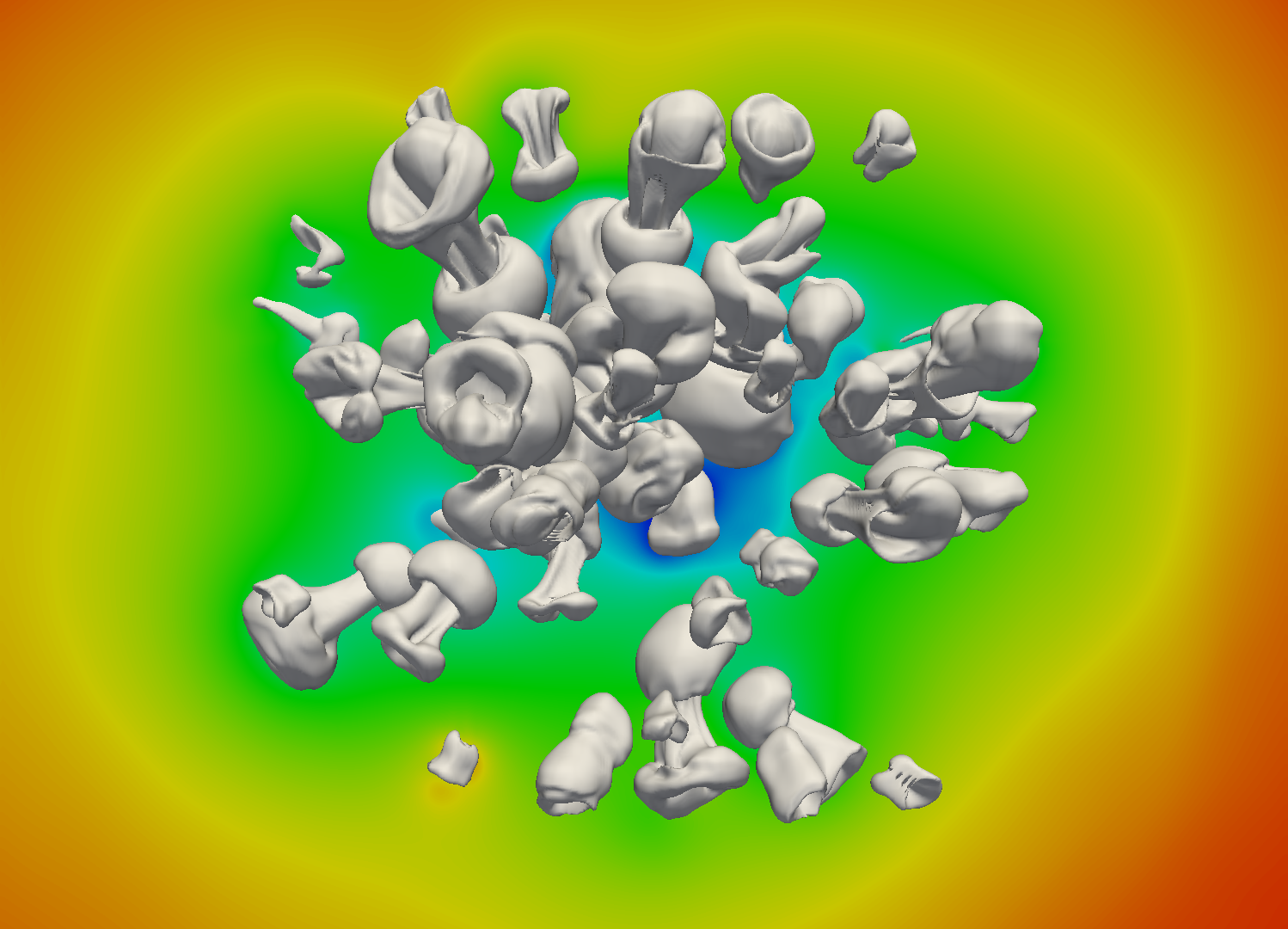}}
      \caption{$10$k steps}
      \label{fig:sim_10k_steps}
    \end{subfigure}

    \caption{Initial bubble cloud and snapshots of cavitating cloud after $5$k and $10$k simulation steps.}
    \label{fig:cloud}
  \end{figure*}

  We evaluate the performance of our compression module after 5k and 10k simulations steps for the following quantities of interest (QoIs): pressure $p$, density $\rho$, total energy $E$ and gas volume fraction $\alpha_2$.
  A statistical description of the datasets is provided in Table~\ref{tab:datasets} while images of the cloud at those steps are depicted in Figure~\ref{fig:cloud}.

  \begin{table*}[!t]
    \caption{Statistics of the QoIs for the cloud cavitation data after 5k (top) and 10k (bottom) simulation steps}
    \centering
    \scriptsize

    \begin{minipage}{1.0\textwidth}
      \label{tab:datasets}
      \centering
      \scriptsize
      \begin{tabular}{| c | r | r | r | r |}
        \hline
        \textbf{QoI}\Top\Bot & \textbf{Min\phantom{3}}  & \textbf{Max\phantom{2}}  & \textbf{Mean\phantom{2}}  & \textbf{StDev\phantom{1}} \\
        \hline
        $\bm{p}$\Top& 4.9e+01 & 9.4e+02 & 3.2e+01 & 5.2e+01 \\
        $\bm{\rho}$ & 1.6e+01 & 1.0e+03 & 1.3e+02 & 3.6e+02 \\
        $\bm{E}$  & 1.2e+02 & 8.0e+03 & 1.0e+03 & 2.9e+03 \\
        $\bm{a_2}$\Bot & 0.0e+00 & 1.0e+00 & 5.9e$-$03 & 6.7e$-$02 \\
        \hline
      \end{tabular}
    \end{minipage}

    \begin{minipage}{1.0\textwidth}
      \scriptsize
      \centering
      \begin{tabular}{| c | r | r | r | r |}
        \hline
        \textbf{QoI}\Top\Bot & \textbf{Min\phantom{3}}  & \textbf{Max\phantom{2}}  & \textbf{Mean\phantom{2}}  & \textbf{StDev\phantom{1}} \\
        \hline
        $\bm{p}$\Top& 1.1e+01  & 1.3e+02  & 1.6e+01  & 2.3e+01 \\
        $\bm{\rho}$ & 1.3e+01  & 1.0e+03  & 1.3e+02  & 3.6e+02 \\
        $\bm{E}$  & 9.4e+01  & 7.8e+03  & 1.0e+03  & 2.9e+03  \\
        $\bm{a_2}$\Bot & 0.0e+00  & 1.0e+00 & 1.5e$-$02 & 9.9e$-$02 \\
        \hline
      \end{tabular}
    \end{minipage}
  \end{table*}

\begin{figure*}[!t]
    \centering
    \begin{subfigure}[t]{1.00\textwidth}
       \resizebox{\textwidth}{!}{\includegraphics{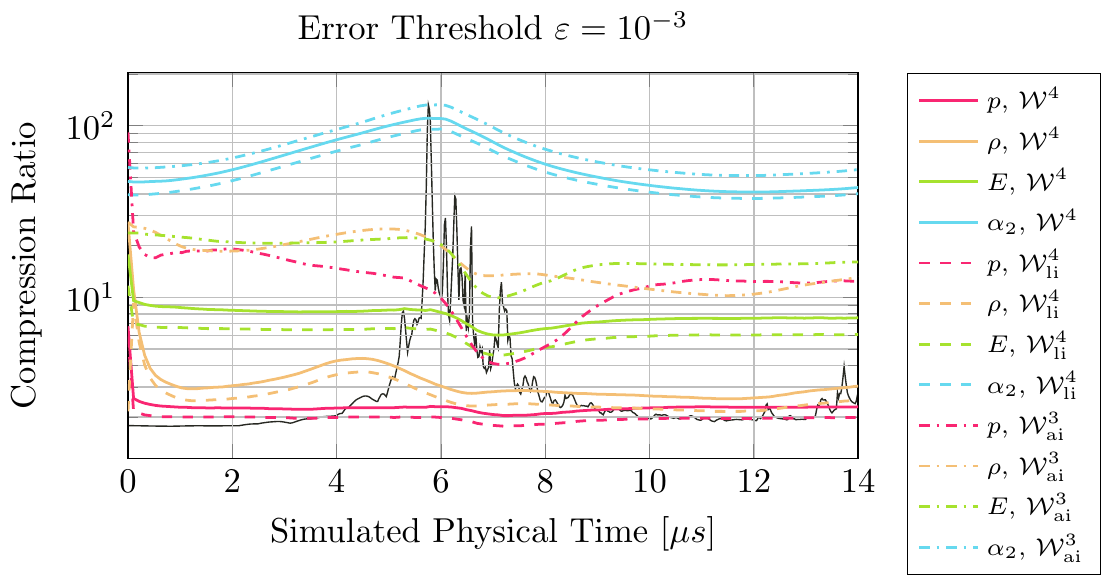}}
    \end{subfigure}
    \begin{subfigure}[t]{1.00\textwidth}
       \resizebox{\textwidth}{!}{\includegraphics{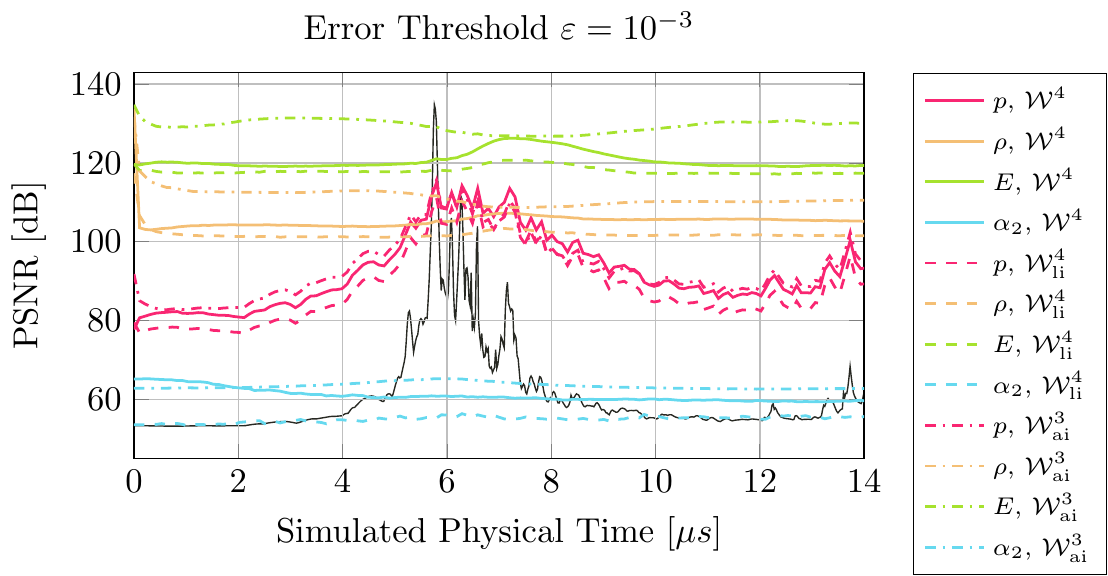}}
    \end{subfigure}
    \caption{Compression rate (top) and peak signal-to-noise-ratio (bottom) over the duration of the cloud collapse simulation for the three wavelet types.
    The thin solid line marks the local peak pressure in the domain and serves as an indicator for strong variation in the data.}
    \label{fig:test_data_time_history}
  \end{figure*}

  The temporal evolution of the compression ratio and the PSNR for the defined QoIs above is shown in Figure~\ref{fig:test_data_time_history} for an error tolerance of $\varepsilon = 10^{-3}$ and the three different wavelet types, that is fourth order wavelets (\wavfourth), fourth order lifted interpolating  (\wavfourthlift) and third order average interpolating (\wavthirdavg) wavelets.
  Furthermore, the figure shows the local peak pressure (thin solid line) which can be used as an indicator for the distortion in the dataset.
  The gas volume fraction $\alpha_2$ describes the topology of the liquid-gas interfaces which are only affected locally by incoming disturbances.
  Due to physical compression of the gas bubbles prior to the collapse of the cloud, the bubbles shrink in size which results in a significant increase of the compression ratio.
  After the collapse of the cloud, there is a rebound phase where bubbles expand again.
  Consequently, the observed compression ratio for this scalar field decreases in the post-collapse phase, c.f.\ Figure~\ref{fig:sim_5k_steps} and~\ref{fig:sim_10k_steps}.
  Finally, the use of the \wavthirdavg{} wavelets results in a significant improvement in the observed compression ratio at low error thresholds for scalar fields that are difficult to compress, such as pressure $p$, density $\rho$ as well as total energy $E$.
  The pressure field is characterized by the largest discontinuities compared to $\rho$ and $E$.  These discontinuities propagate through the computational domain within a short amount of time due to the high wavespeeds associated with them.
  The peak of the collapse happens around $t=7\,\mu s$.  Shortly after that time instant, strong discontinuities propagate outward from the center of the domain.
  During that time, a reduction of the compression ratio is observed for all of these fields using the \wavthirdavg{} wavelets, where the reduction is most significant in the pressure field due to the emission of strong shock waves.
  The \wavthirdavg{} wavelets evidently outperform the other wavelet types, as can further be seen from Figure~\ref{fig:test_data_time_history}.

  \subsection{Wavelet-based data compression}

\begin{figure*}[!ht]
    \centering
    \begin{subfigure}[t]{0.45\textwidth}
      \centering
      \resizebox{\textwidth}{!}{\includegraphics{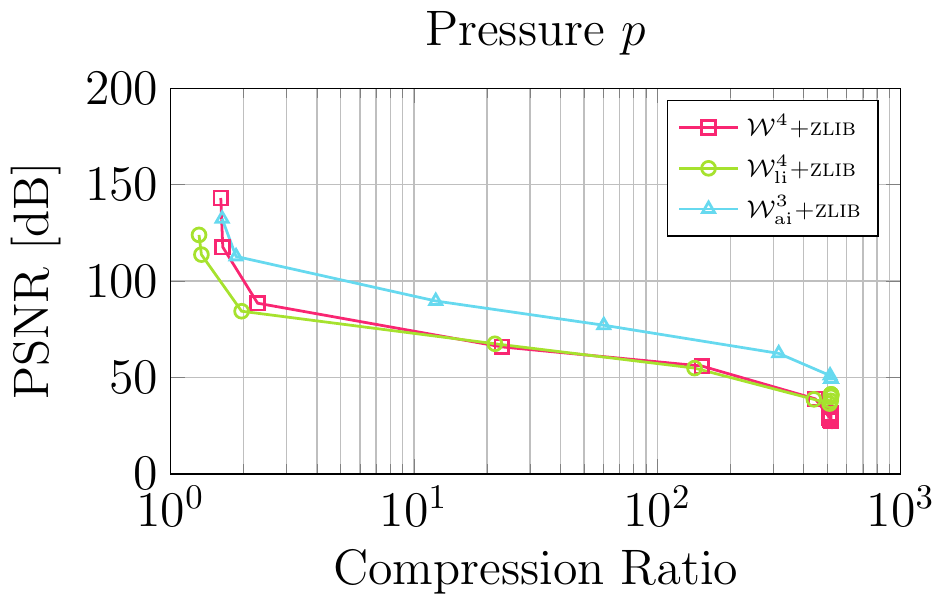}}
    \end{subfigure}%
    \hspace{1em}
    \begin{subfigure}[t]{0.45\textwidth}
      \centering
      \resizebox{\textwidth}{!}{\includegraphics{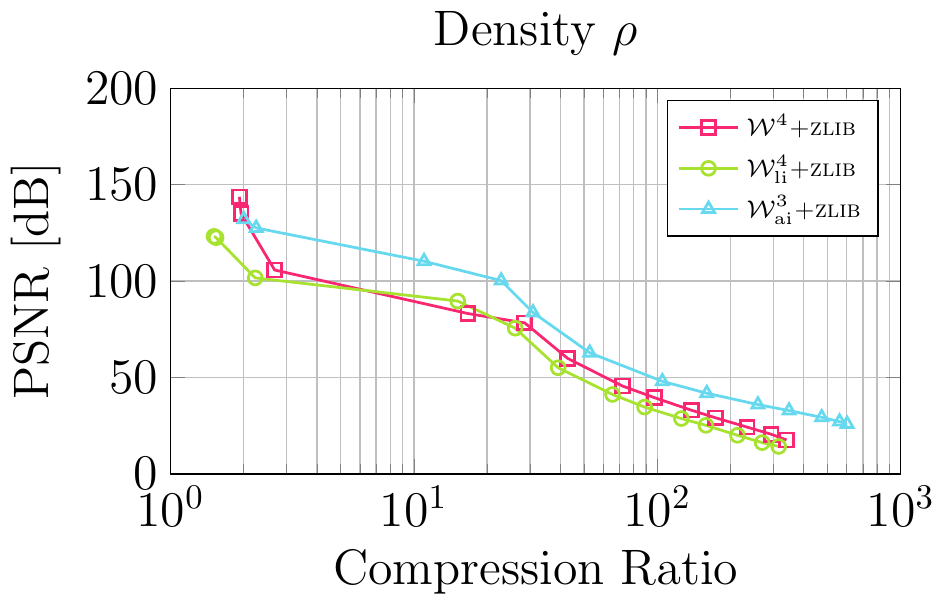}}
    \end{subfigure}
    \caption{Compression performance for three different types of wavelets and
    two QoIs ($p$, $\rho$) after 10k simulation steps.}
    \label{fig:compression_exp1}

    \centering
    \begin{subfigure}[t]{0.45\textwidth}
      \centering
      \resizebox{\textwidth}{!}{\includegraphics{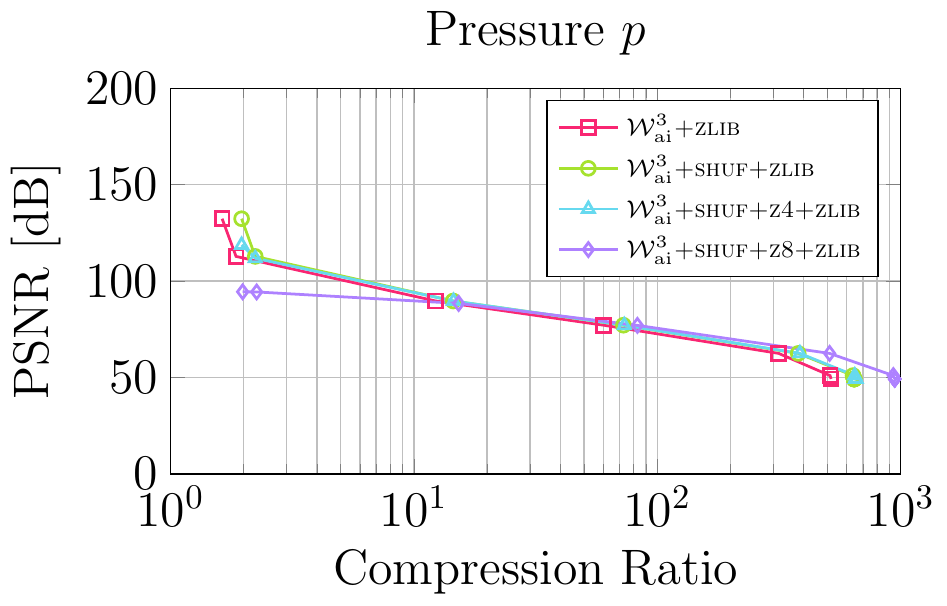}}
    \end{subfigure}%
    \hspace{1em}
    \begin{subfigure}[t]{0.45\textwidth}
      \centering
      \resizebox{\textwidth}{!}{\includegraphics{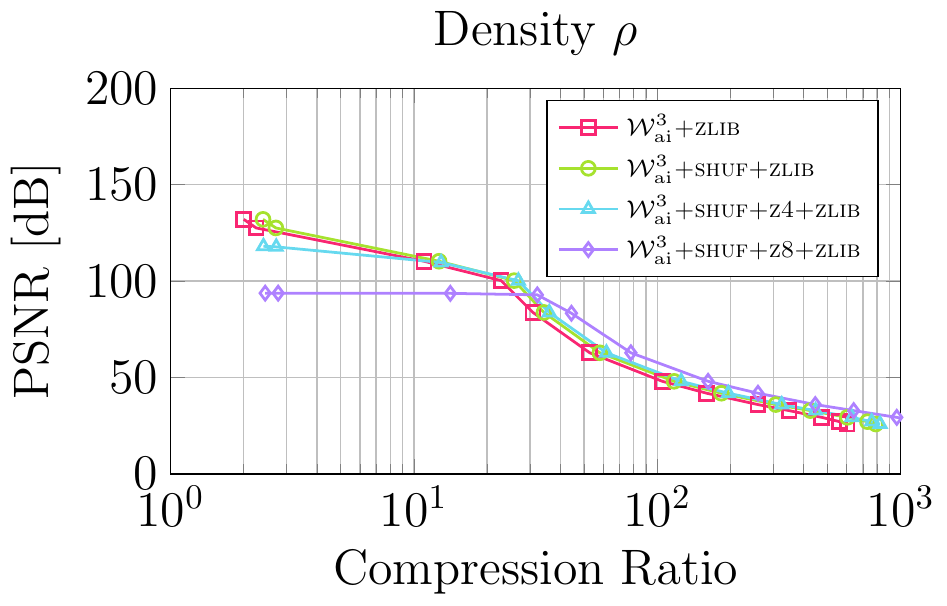}}
    \end{subfigure}

    \caption{Compression performance for the third order wavelets combined with data (byte) shuffling and bit zeroing and two QoIs ($p$, $\rho$) after 10k simulation steps.}
    \label{fig:compression_exp2}

    \centering
    \begin{subfigure}[t]{0.45\textwidth}
      \centering
      \resizebox{\textwidth}{!}{\includegraphics{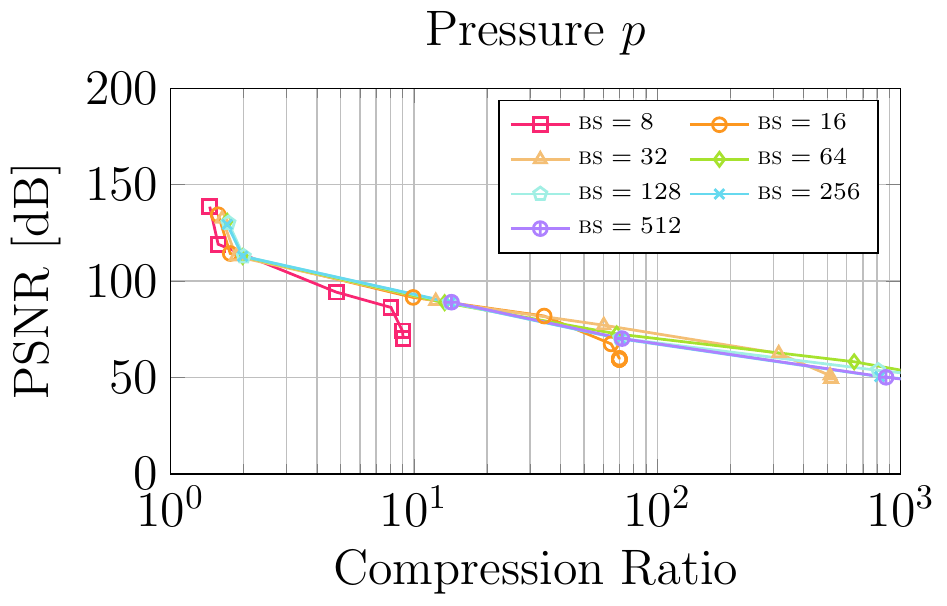}}
    \end{subfigure}%
    \hspace{1em}
    \begin{subfigure}[t]{0.45\textwidth}
      \centering
      \resizebox{\textwidth}{!}{\includegraphics{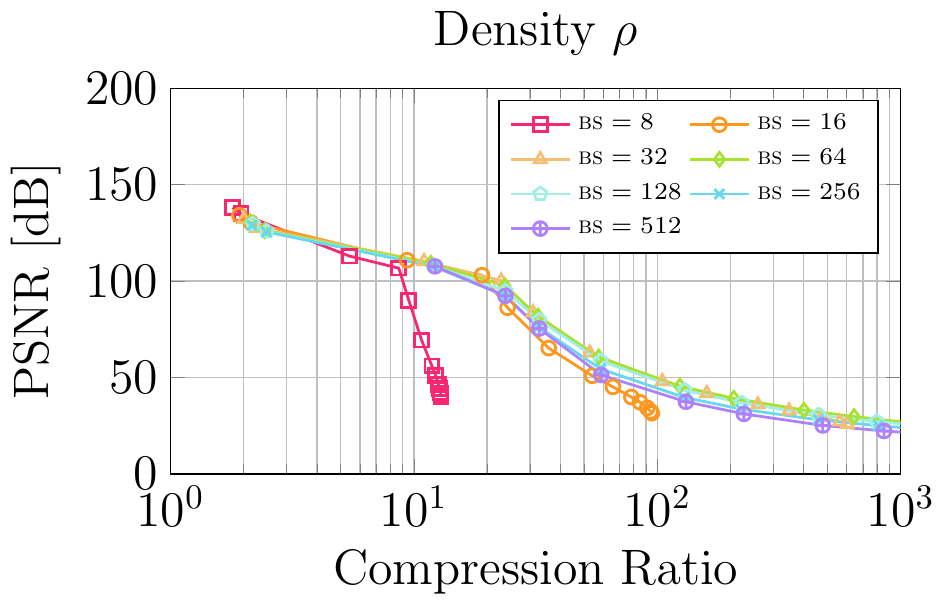}}
    \end{subfigure}

    \caption{Compression performance for various block sizes and two QoIs ($p$, $\rho$) after 10k simulation steps.}
    \label{fig:compression_exp3}
  \end{figure*}

  We perform the following experiments:

  \paragraph{Exp.\ 1} Using ZLIB at its default compression level as our encoder, we compare the three available wavelet types for two fields (pressure $p$ and density $\rho$).
  In Figure~\ref{fig:test_data_time_history} and~\ref{fig:compression_exp1}, we observe that the third-order wavelets (\wavthirdavg) outperform the other two types, which show comparable behavior.

  \paragraph{Exp.\ 2} For the best wavelet type (\wavthirdavg), we study the effect of bit zeroing and data shuffling (Figure~\ref{fig:compression_exp2}).
  $Z4/Z8$ denote that we zero the 4/8 least significant bits of the wavelet detail coefficients, respectively.
  Data shuffling is performed on the aggregate buffer at byte level with block size equal to 4 bytes, in accordance to the single precision floating point data.
  We see that data shuffling improves compression ratio without affecting PSNR.
  This is expected as all data shuffling operations are reversible, while higher compression ratio can be achieved if the extracted bit planes are ``boring''.
  Moreover, bit zeroing can be beneficial and increase compression ratio for values of PSNR lower that a threshold imposed by the accuracy loss
  (e.g.\ flatness region for $Z8$ in Figure~\ref{fig:compression_exp2}).

  Shuffling the aggregate buffer instead of only the detail coefficients provides $\approx$4\% higher compression.
  For the same type of wavelets and datasets, LZMA results in $\approx$14\% higher compression ratio than ZLIB.
  This improvement, however, becomes less than 3\% when data shuffling has been previously applied to the detail coefficients.

  \paragraph{Exp.\ 3} We study the effect of block size, a configuration option that is strongly related to the performance tuning of the simulation software.
  Figure~\ref{fig:compression_exp3} shows that the compression ratio is significantly lower for small block sizes ($8^3$ and $16^3$), while larger block sizes exhibit similar behavior.
  For our simulations on IBM BG/Q systems, blocks of $32^3$ and $64^3$ points represent the best options, in terms of simulation time~\cite{Hadjidoukas:2015a} and compression ratio.
  This is attributed to the optimal memory footprint of a single block and the amount of work determined by the total number of blocks and exploited by thread-level parallelism. \\

  \subsection{Comparison of data compression methods}

  \begin{figure*}[!t]
    \centering
    \begin{subfigure}[t]{0.44\textwidth}
      \centering
      \resizebox{\textwidth}{!}{\includegraphics{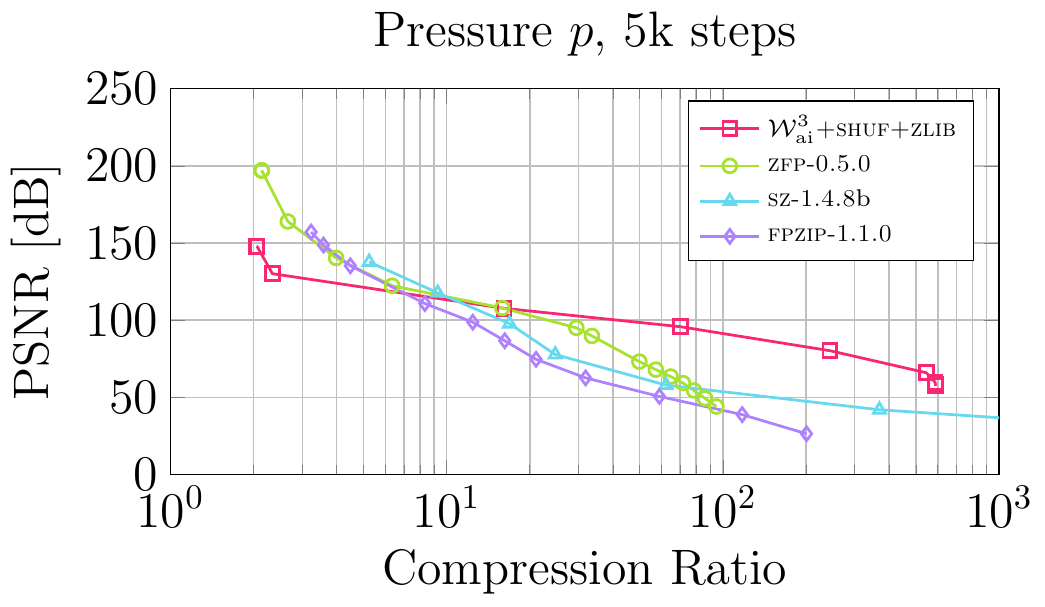}}
    \end{subfigure}%
    \hspace{1em}
    \begin{subfigure}[t]{0.44\textwidth}
      \centering
      \resizebox{\textwidth}{!}{\includegraphics{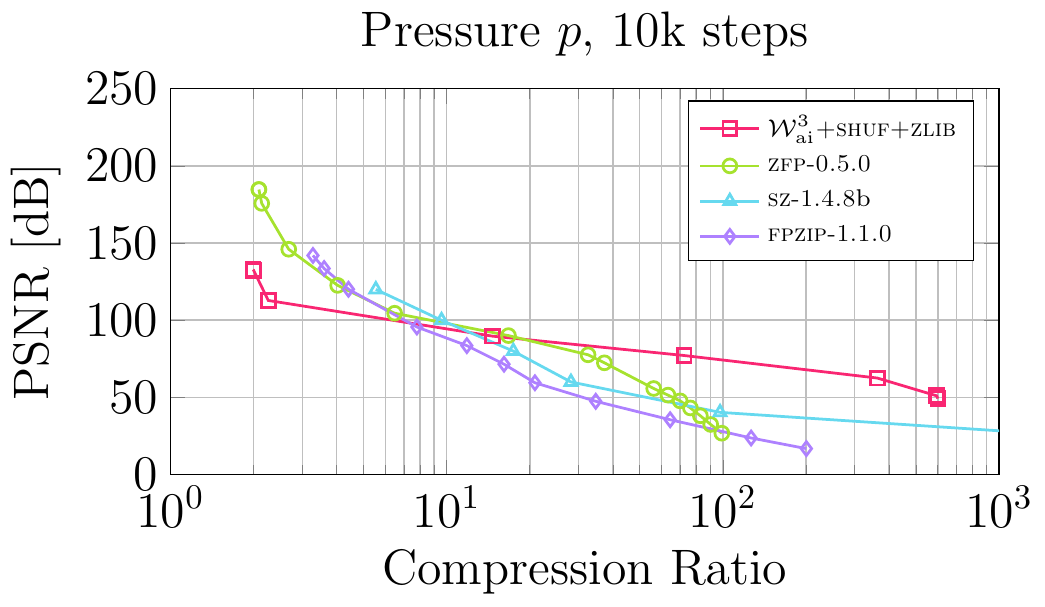}}
    \end{subfigure}

    \begin{subfigure}[t]{0.44\textwidth}
      \centering
      \resizebox{\textwidth}{!}{\includegraphics{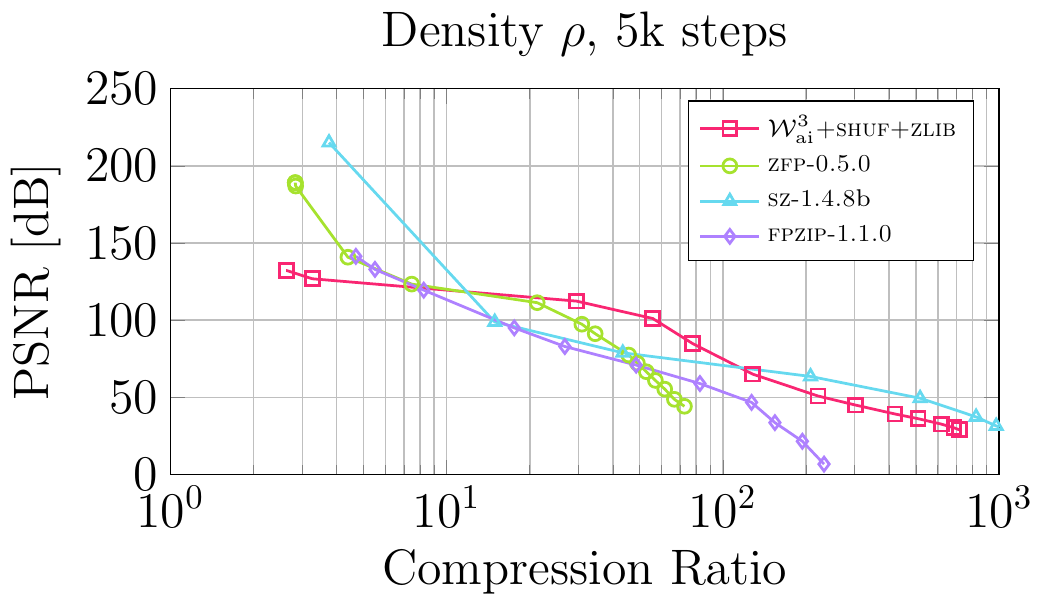}}
    \end{subfigure}%
    \hspace{1em}
    \begin{subfigure}[t]{0.44\textwidth}
      \centering
      \resizebox{\textwidth}{!}{\includegraphics{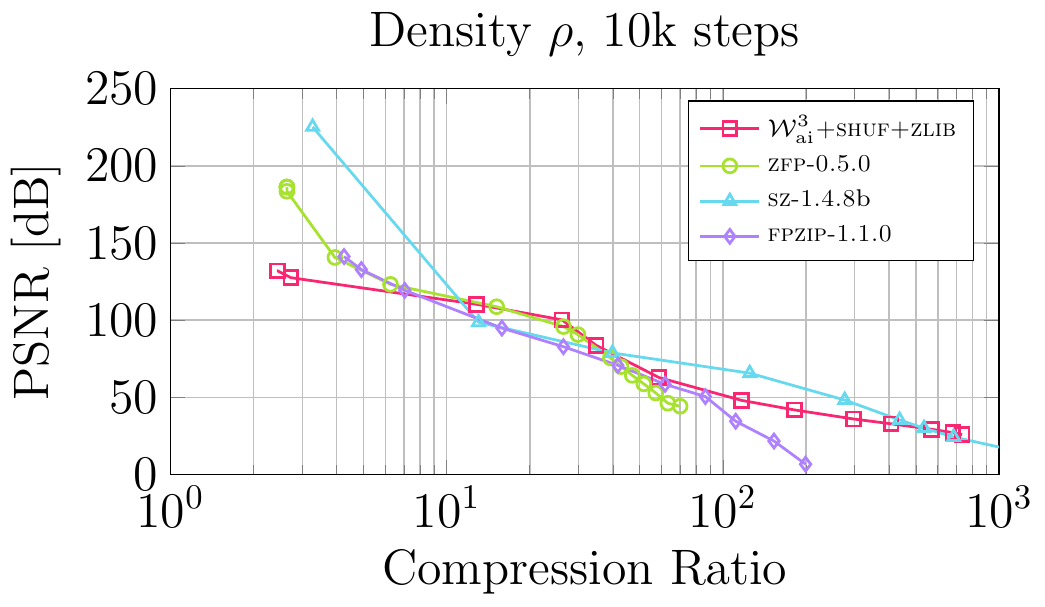}}
    \end{subfigure}

    \begin{subfigure}[t]{0.44\textwidth}
      \centering
      \resizebox{\textwidth}{!}{\includegraphics{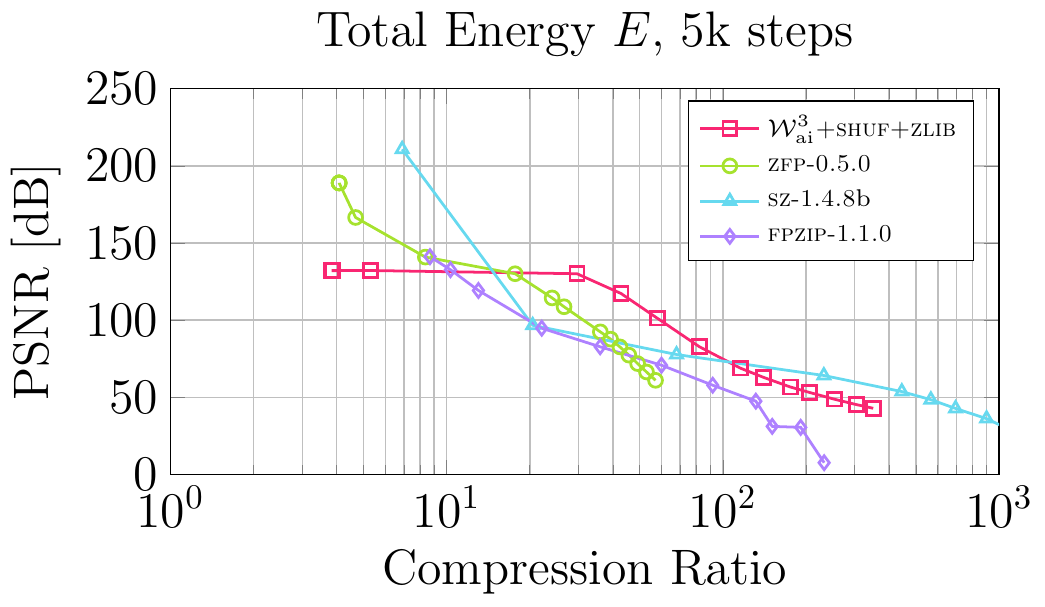}}
    \end{subfigure}%
    \hspace{1em}
    \begin{subfigure}[t]{0.44\textwidth}
      \centering
      \resizebox{\textwidth}{!}{\includegraphics{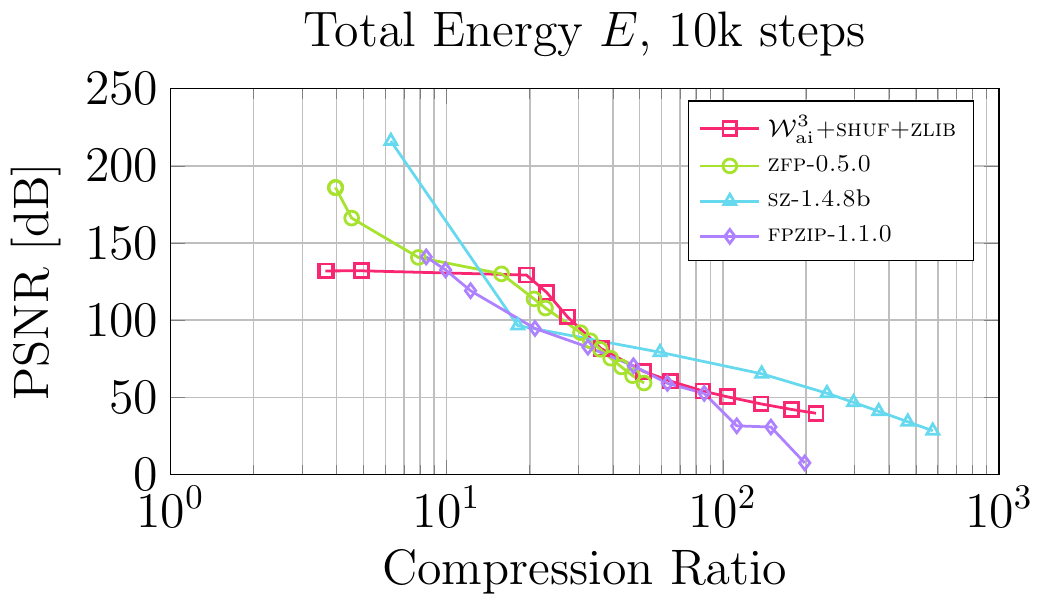}}
    \end{subfigure}

    \begin{subfigure}[t]{0.44\textwidth}
      \centering
      \resizebox{\textwidth}{!}{\includegraphics{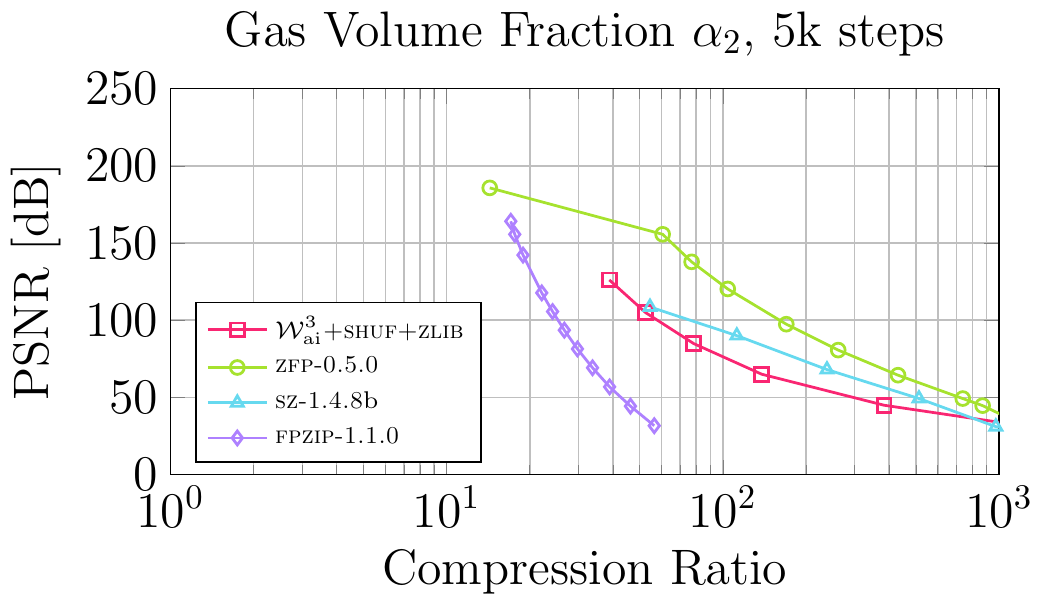}}
    \end{subfigure}%
    \hspace{1em}
    \begin{subfigure}[t]{0.44\textwidth}
      \centering
      \resizebox{\textwidth}{!}{\includegraphics{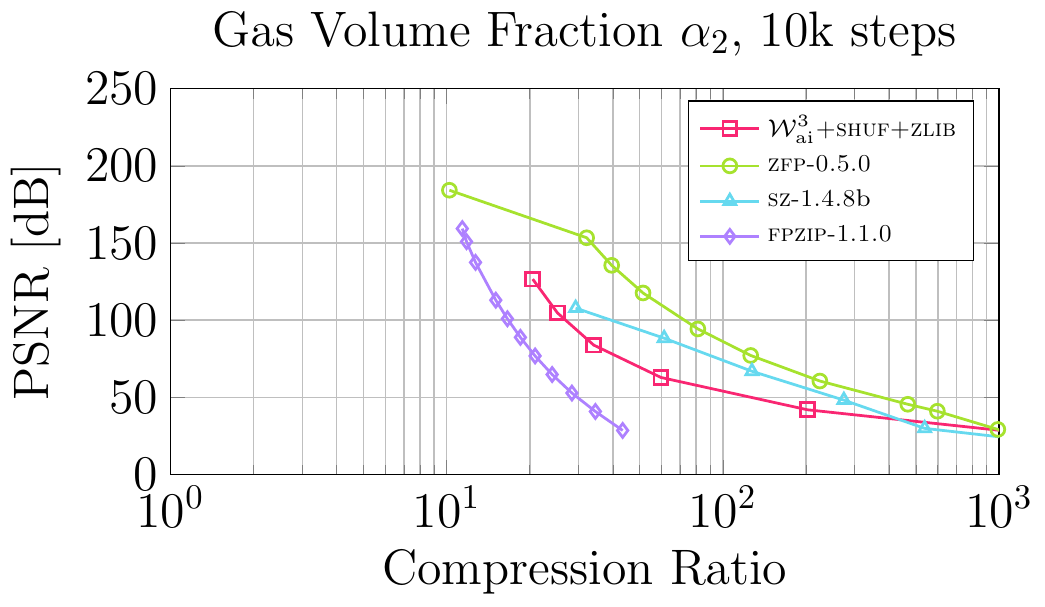}}
    \end{subfigure}

    \caption{PSNR vs compression ratio for $p$, $\rho$, $E$ and $\alpha_2$ 
    after 5k and 10k timesteps.} 
    \label{fig:comparison}
  \end{figure*}

  \begin{figure*}[!t]
    \centering
    \begin{subfigure}[t]{0.44\textwidth}
      \centering
      \resizebox{\textwidth}{!}{\includegraphics{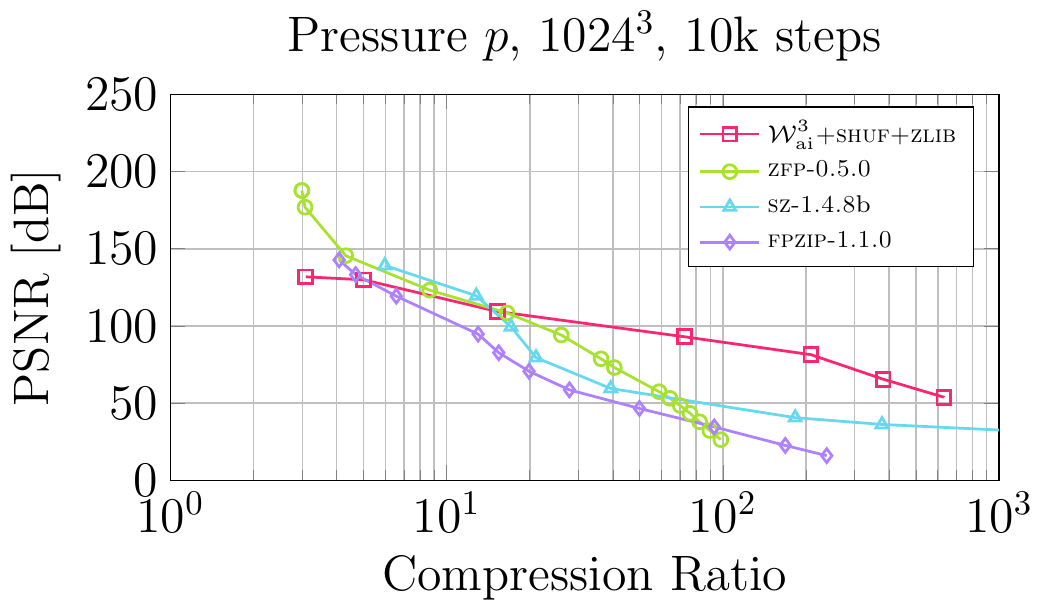}}
    \end{subfigure}%
    \hspace{1em}
    \begin{subfigure}[t]{0.44\textwidth}
      \centering
      \resizebox{\textwidth}{!}{\includegraphics{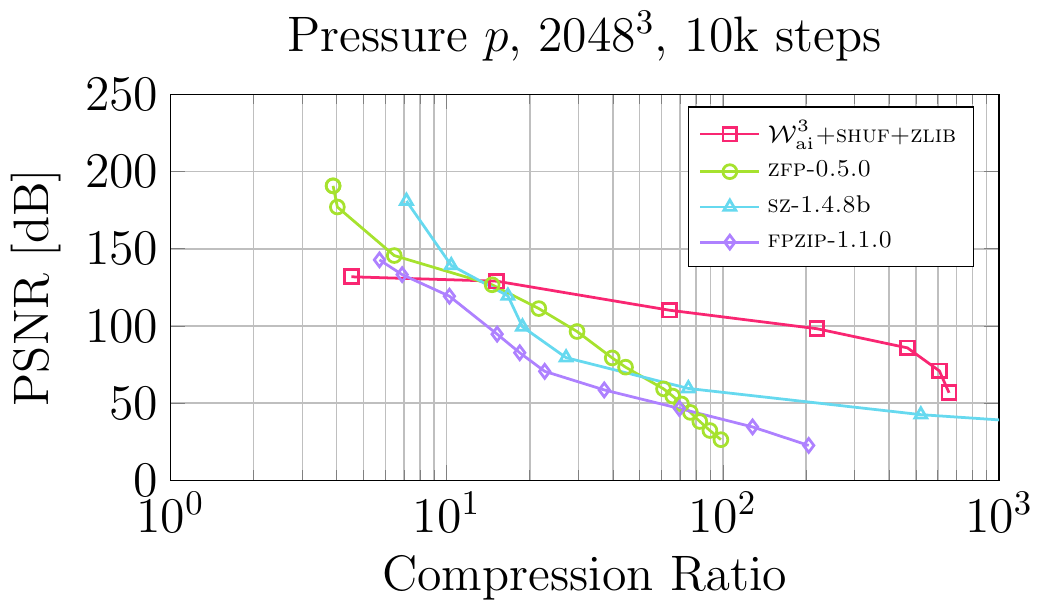}}
    \end{subfigure}

    \begin{subfigure}[t]{0.44\textwidth}
      \centering
      \resizebox{\textwidth}{!}{\includegraphics{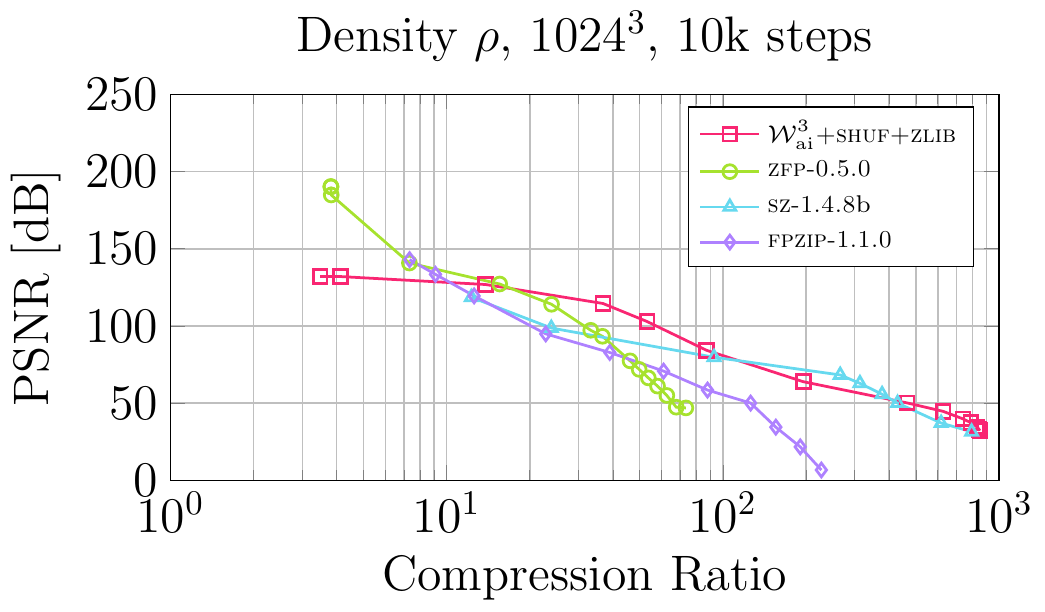}}
    \end{subfigure}
    \hspace{1em}
    \begin{subfigure}[t]{0.44\textwidth}
      \centering
      \resizebox{\textwidth}{!}{\includegraphics{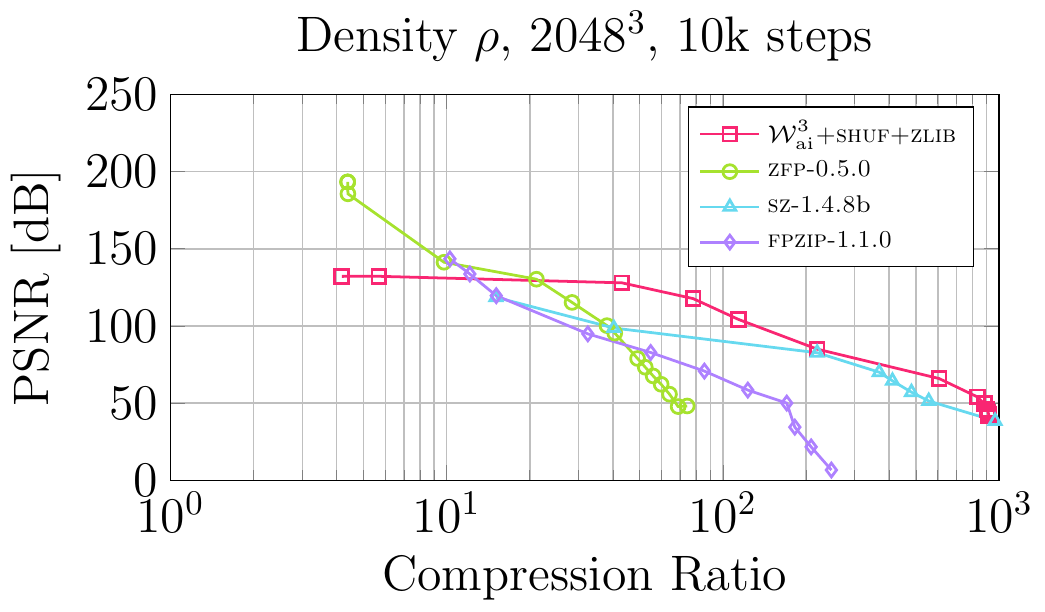}}
    \end{subfigure}

    \begin{subfigure}[t]{0.44\textwidth}
      \centering
      \resizebox{\textwidth}{!}{\includegraphics{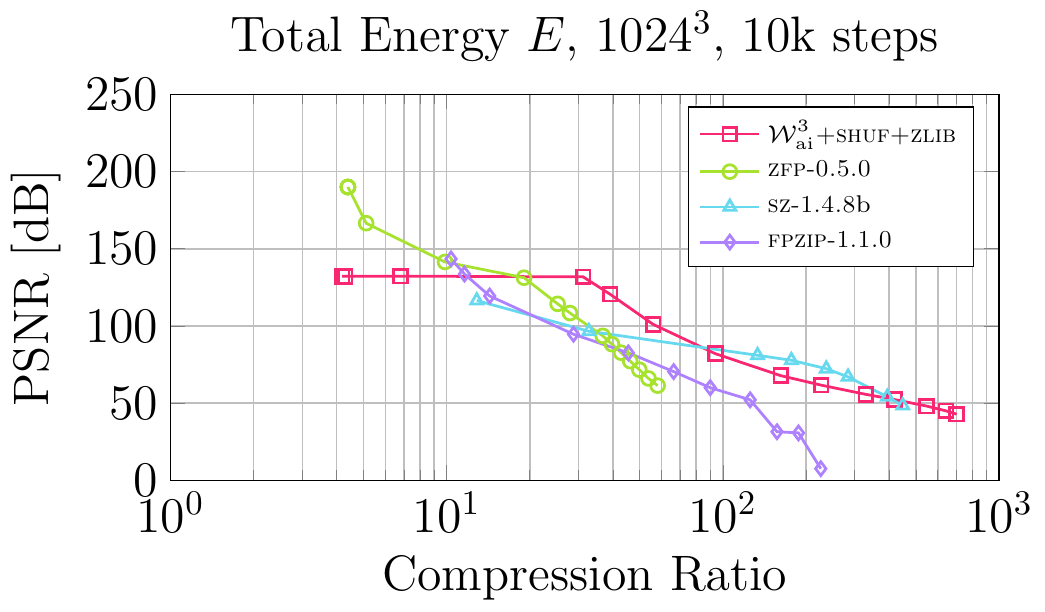}}
    \end{subfigure}%
    \hspace{1em}
    \begin{subfigure}[t]{0.44\textwidth}
      \centering
      \resizebox{\textwidth}{!}{\includegraphics{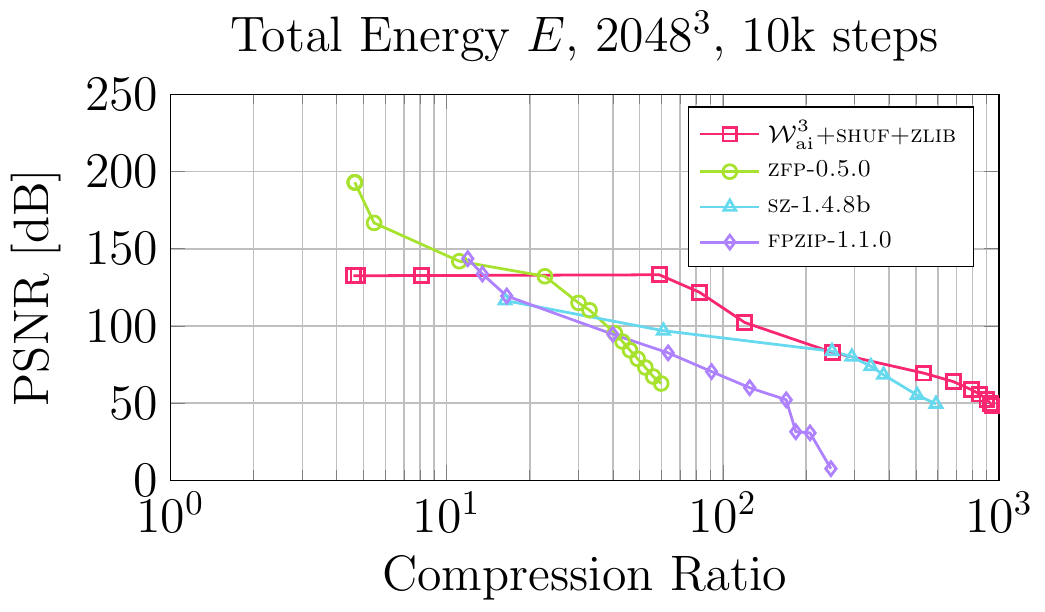}}
    \end{subfigure}

    \begin{subfigure}[t]{0.44\textwidth}
      \centering
      \resizebox{\textwidth}{!}{\includegraphics{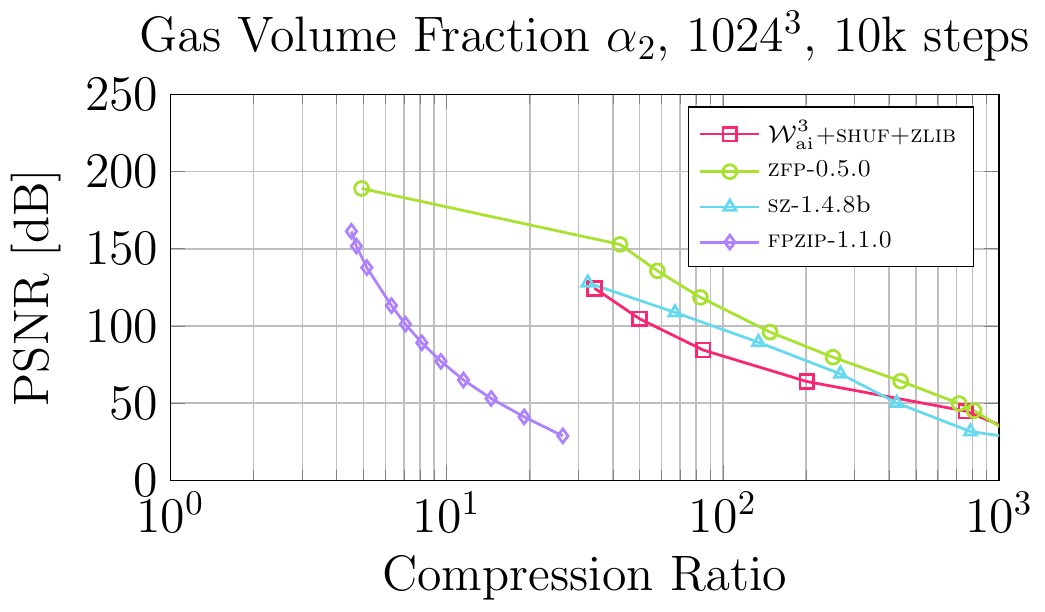}}
    \end{subfigure}
    \hspace{1em}
    \begin{subfigure}[t]{0.44\textwidth}
      \centering
      \resizebox{\textwidth}{!}{\includegraphics{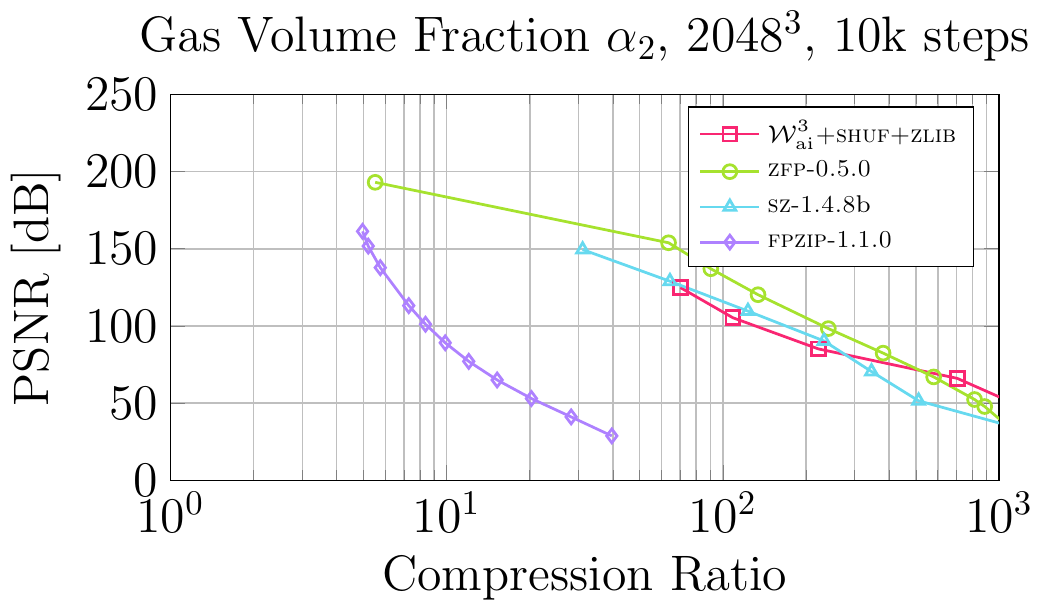}}
    \end{subfigure}

    \caption{PSNR vs compression ratio for $p$, $\rho$, $E$ and $\alpha_2$
    after 10k timesteps, for $1024^3$ (left) and $2048^3$ (left) grid points.}
    \label{fig:comparison2}
  \end{figure*}

  We evaluate the following lossy compression methods available in CubismZ: wavelets, ZFP (v0.5.0), SZ (v1.4.11) and FPZIP (v1.1.0).
  Based on the previously presented results, we use third order average interpolating wavelets, combined with byte shuffling and lossless compression using ZLIB.

  We perform our experiments by varying the error threshold for wavelets and ZFP, as well as the relative error bound for SZ, so as to identify the range of PSNR values that can be achieved by each method.
  Similarly, we vary the bits of precision of the FPZIP compressor. Any additional configuration options, for instance for SZ, are set to their default values.
  A representative comparison regarding the compression ratio and quality of the four methods for 3D cloud cavitation simulation data is depicted in Figure~\ref{fig:comparison}.
  We observe that there is no method that outperforms the others in all cases. Instead, the compression performance varies with the dataset and the requested accuracy, i.e., the PSNR value which is related to the subsequent usage of the data dumps.
  For the first three fields ($p$, $\rho$ and $E$), wavelets are more suitable for our visualization needs as they yield highest compression ratios for PSNR values in the range of 100-120 dB.
  Moreover, they outperform or compete with SZ for lower values of PSNR.
  For high PSNR values, both ZFP and SZ achieve the best compression performance, while ZFP clearly outperforms the other methods in the compression of the gas volume fraction $\alpha_2$.
  ZFP exhibits better performance than SZ in several cases, which is attributed to the fact that ZFP is more suitable for 3D data~\cite{Di:2016}.
  We also observe that the obtained compression ratios are lower for the datasets generated after 10k timesteps, because we are closer to the local peak pressure.

  Figure~\ref{fig:comparison2} depicts PSNR vs.\ compression ratio plots for datasets produced by the same simulation running at higher resolution.
  More specifically, we show  results for $1024^3$ and $2048^3$ grid points and 10k timesteps, complementing those depicted in Figure~\ref{fig:comparison} for $512^3$ points.
  We observe that higher resolution improves the performance of the wavelet-based compression, in contrast to the other three methods which yield similar results regardless of the number of grid points.

  \paragraph{Floating point compression of the wavelet coefficients}
  In this experiment, we use third-order wavelets (\wavthirdavg) and a specific error threshold in the first substage of CubismZ  in order to study the compression performance when the wavelet detail coefficients are processed by the FPZIP, SZ and SPDP floating-point compressors.
  In contrast to the other compressors, SPDP~\cite{SPDP:2017} is tailored to sequences of single and double-precision floating-point data and achieves higher lossless compression than the double-precision FPC~\cite{Burtscher:2009} compressor.

  As before, the aggregate buffer with the compressed coefficients and bit-set masks are further compressed with ZLIB.
  The PSNR value is determined by the first substage and is unaffected by the subsequent lossless compression techniques.
  Table~\ref{tab:exp4} shows the achieved compression ratios for three tolerance levels and PSNR values for one of our input datasets ($p$ after 10k simulation steps).
  We observe that SPDP outperforms both FPZIP and SZ but none of the compressors provide better overall compression ratio than byte shuffling combined with ZLIB. Byte and bit shuffling do not improve the compression ratio when the aggregate buffer already contains compressed wavelet coefficients.

    \begin{table}[h]
    \caption{Compression ratio values for the third order wavelets combined with various techniques applied to the wavelet coefficients.
    The private buffer undergoes lossless compression with ZLIB.
    The input dataset is $p$ after 10k simulation steps using three values of error tolerance $\varepsilon$.}

    \label{tab:exp4}
    \centering
    \scriptsize
    \begin{tabular}{| l || c | c | c |}
      \hline
      \multirow{2}{*}{\textbf{Wavelets}} & $\varepsilon=10^{-4}$\Top & $\varepsilon=10^{-3}$ & $\varepsilon=10^{-2}$ \\
                                         & 112.8 dB\Bot & 89.7 dB & 77.1 dB \\
      \hline
      \hline
      +FPZIP+ZLIB\Top     & 1.26    &  \phantom{1}9.81  & 56.00 \\
      +SZ+ZLIB               & 1.71   & 11.65 & 59.54 \\
      +SPDP+ZLIB\Bot      & 1.98   & 12.93 & 64.86 \\
      \hline
      +ZLIB\Top                  & 1.85  & 12.22  & 60.14 \\
      +SHUF+ZLIB\Bot      &  2.26  & 14.64  & 72.36 \\
      \hline
    \end{tabular}

  \end{table}

\section{Performance evaluation}
  \label{sec:performance}

  In this section, we study the parallel performance of CubismZ. As our software targets parallel and distributed systems based on multicores, we performed our experiments on the Piz Daint (CSCS) and MIRA IBM BG/Q (ANL) supercomputers.

  \subsection{Single-core performance}

  We evaluate the sequential performance for our wavelet-based scheme on a single node of Piz Daint.
  The node is equipped with a 12-core Intel Xeon E5-2690v3 processor and 64GB memory.
  Table \ref{tab:speeds} shows the compression and decompression speeds (in MB/s) of various schemes, for a specific dataset (pressure $p$ at 10k simulation steps) and PSNR value.
  We set the parameter (error tolerance or bound) of the lossy compressors so as to achieve similar PSNR value ($\approx$90 dB).

We observe that the compression speed of the wavelet-based scheme strongly depends on the performance of the lossless compressor.
Byte shuffling increases both compression ratio and speed and when combined with ZSTD yields similar performance to that of SZ.
For the specific experiment, ZFP exhibit better compression and decompression speedups.
However, neither our third-order wavelets nor SZ have been optimized in performance~\cite{Di:2016}.
The third group of measurements showcases the performance of lossless compression and how this compares to lossy compression.

    \begin{table}[h]
    \caption{Compression ratio and compression and decompression speeds (MB/s) for various schemes.
     The input dataset is pressure $p$ after 10k simulation steps and the resulted PSNR value is similar for all lossy compressors.}

    \label{tab:speeds}
    \centering
    \scriptsize
    \begin{tabular}{| c | c | r | C{1cm} | C{1cm} |}
      \hline
      Stage1\Top\Bot               & Stage2               & CR\phantom{-}      & Comp               & Decomp \\
      \hline
      \hline
      \multirow{5}{*}{\wavthirdavg}  &  -\Top          &   7.71 &            277              &  215  \\
                                         & ZLIB                        & 12.22 &             50              &   184   \\
                                         & SHUF+ZLIB             & 14.64 &             66              &   178   \\
                                         & SHUF+ZSTD           & 14.47 &             89              &   204  \\
                                         & SHUF+LZ4HC\Bot   & 13.03 &           131              &   212   \\
      \hline
      ZFP                            &   -\Top                       & 16.74  &          126             &  507  \\
      SZ                              &    -                             & 14.45 &            62            &    109  \\
      FPZIP                         &     -\Bot                    &  9.64   &          119            &    99  \\
      \hline
      SHUF+ZLIB                &      -\Top                  & 2.00   &            51             &  322  \\
      SHUF+ZSTD              &       -\Bot                 &  1.98  &            88             &  883  \\
      \hline
    \end{tabular}
  \end{table}

  Table~\ref{tab:zlib} shows PSNR, compression ratios and execution time on a single core for the wavelet-based \wavthirdavg+ZLIB scheme and three tolerance levels $\varepsilon\in\{10^{-4},10^{-3},10^{-2}\}$.
  The timing measurements include the required data transformations and substages of the data compression scheme but not the time for writing the compressed data to disk.
  We show results for ZLIB and its default compression level (Z/DEF), because of its good balance between speedup and compression ratio and its previous usage in all our large scale production simulations of cloud cavitation.
  We also study the effect of the highest compression level of ZLIB, denoted as Z/BEST, in the compression time.

  We observe that the compression time strongly depends on the tolerance, i.e., the number of detail coefficients passed to ZLIB.
  Z/BEST increases significantly the runtime for negligible improvement of the compression ratio.
  The use of fast lossless compression (such as Z/DEF) is beneficial, especially when the number of floating point wavelet coefficients increases for smaller error tolerance.
  The corresponding single-threaded compression times (T1) range from 3.4 to 5.0 seconds for ZFP and from 8.3 to 9.6 seconds for SZ, for similar PSNR values (in increasing order).
   Wavelets yield better performance than the other two methods, both in terms of compression ratio and execution time as the error tolerance increases and the PSNR value decreases accordingly.

 \begin{table}[t]
    \caption{PSNR, compression ratio values and execution time for \wavthirdavg{} wavelets and ZLIB with the default and best compression levels.}
    \label{tab:zlib}
    \centering
    \scriptsize
    \begin{tabular}{| c | r || r | r | r | r |}
      \hline
      \multirow{2}{*}{\textbf{Tolerance}} & \multirow{2}{*}{\textbf{PSNR\phantom{-}}} & \multicolumn{2}{c|}{\textbf{Z/DEF}}\Top & \multicolumn{2}{c|}{\textbf{Z/BEST}} \\
       &  & CR\phantom{-} & $T_1 (s)$     & CR\phantom{-}    & $T_1 (s)$\Bot \\
      \hline
      $\varepsilon = 10^{-4}$\Top & 112.8 dB & 1.85 & 79.4  & 1.86      &  173.1 \\
      $\varepsilon = 10^{-3}$        & 89.7 dB  & 12.22 & 10.2  & 12.30  &    26.8 \\			
      $\varepsilon = 10^{-2}$\Bot & 77.1 dB   & 60.14 & 3.1    & 60.27  &   4.0 \\			
      \hline
    \end{tabular}
  \end{table}

  \subsection{Multicore performance}

\begin{figure}[t]
    \centering
    \begin{subfigure}[t]{0.45\textwidth}
      \centering
      \resizebox{\textwidth}{!}{\includegraphics{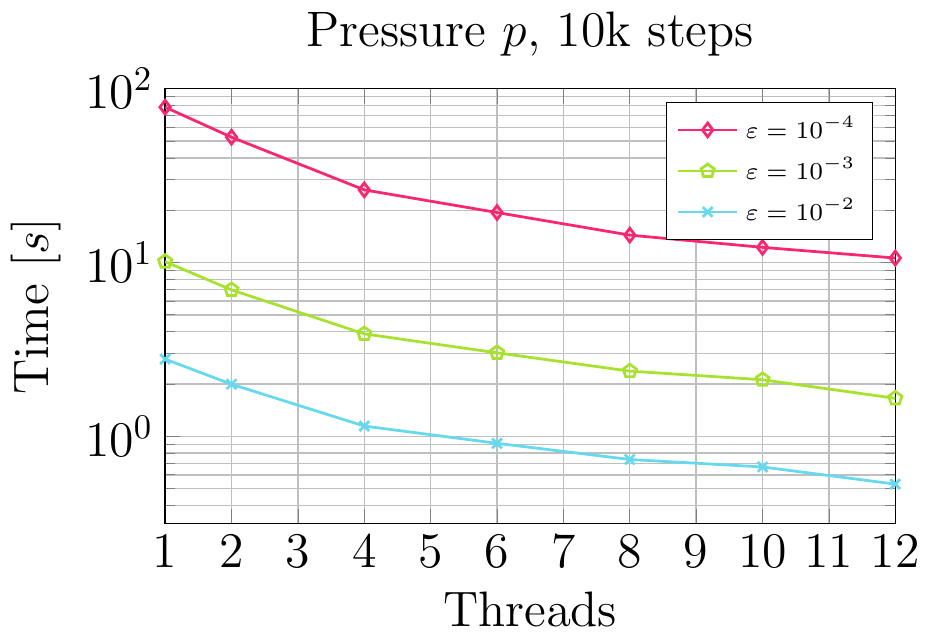}}
    \end{subfigure}%
    \hspace{1em}
    \begin{subfigure}[t]{0.42\textwidth}
      \centering
      \resizebox{\textwidth}{!}{\includegraphics{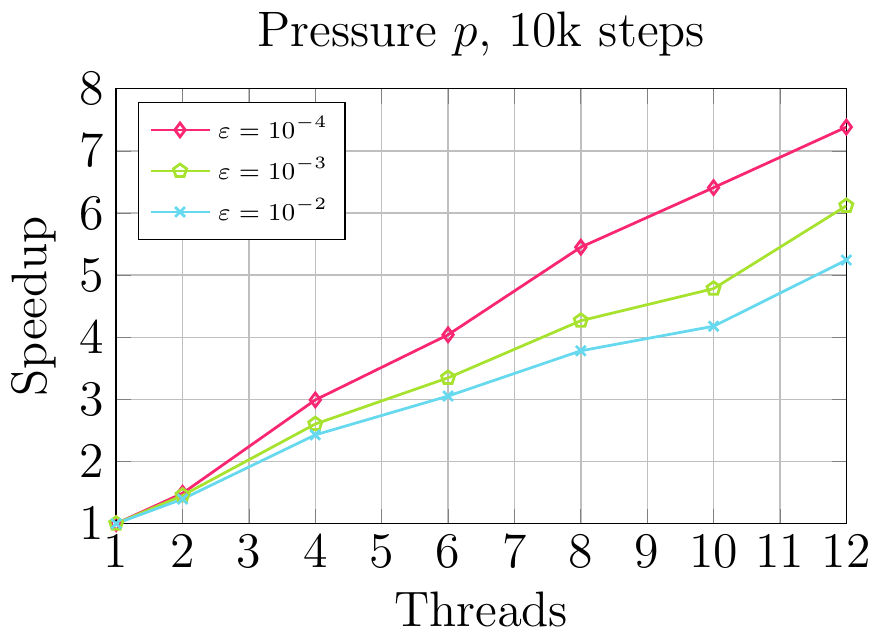}}
    \end{subfigure}

    \begin{subfigure}[t]{0.45\textwidth}
      \centering
      \resizebox{\textwidth}{!}{\includegraphics{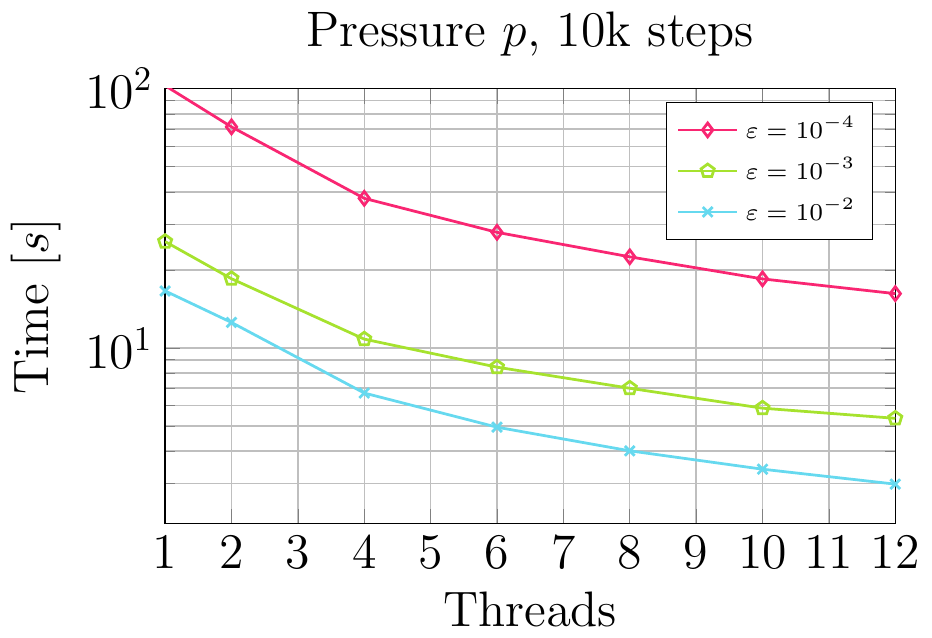}}
    \end{subfigure}%
    \hspace{1em}
    \begin{subfigure}[t]{0.42\textwidth}
      \centering
      \resizebox{\textwidth}{!}{\includegraphics{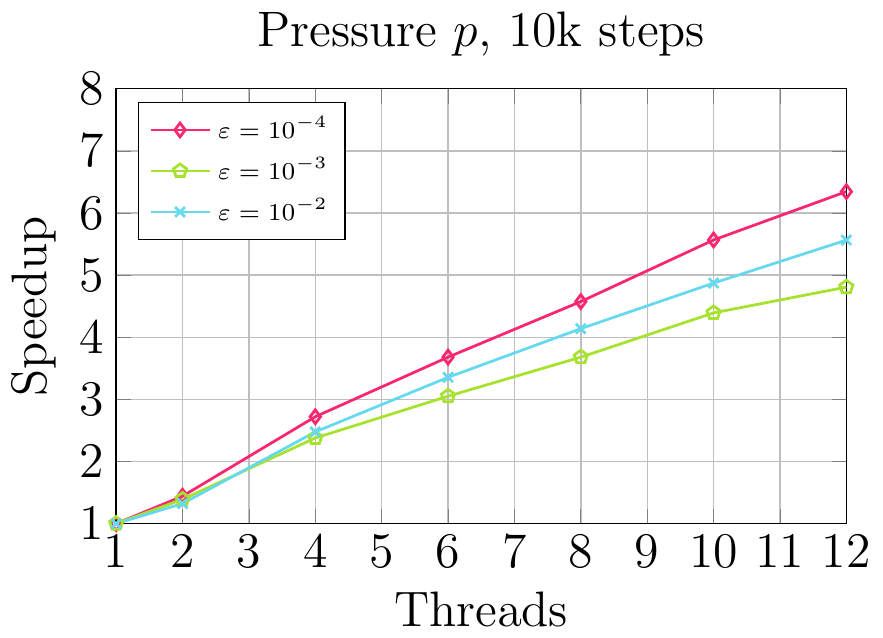}}
    \end{subfigure}

    \caption{Compression time and speedup for wavelets and ZLIB on Piz Daint for the same dataset using $512^3$ (top) and $1024^3$ (bottom) grid points.}
    \label{fig:daint1}
  \end{figure}

  We evaluate the parallel performance of the wavelet-based scheme on a single node of Piz Daint. As before, we use the default compression level of ZLIB.
  Figure~\ref{fig:daint1} illustrates the compression times and corresponding speedups on up to 12 cores for the pressure $p$ dataset using $512^3$ and $1024^3$ grid points (top and bottom, respectively).
  We observe that the code scales better when lossless compression is applied to larger volumes of intermediate data produced by the wavelets at the first compression substage (i.e.\ for $\varepsilon=10^{-4})$.

\begin{figure}[t]
    \centering
    \begin{subfigure}[t]{0.45\textwidth}
      \centering
      \resizebox{\textwidth}{!}{\includegraphics{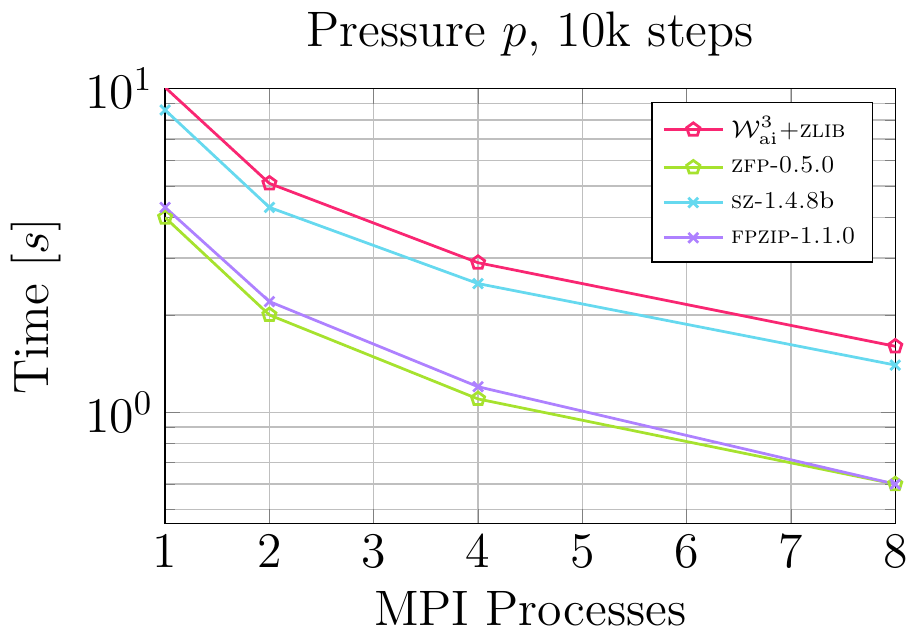}}
    \end{subfigure}%
    \hspace{1em}
    \begin{subfigure}[t]{0.42\textwidth}
      \centering
      \resizebox{\textwidth}{!}{\includegraphics{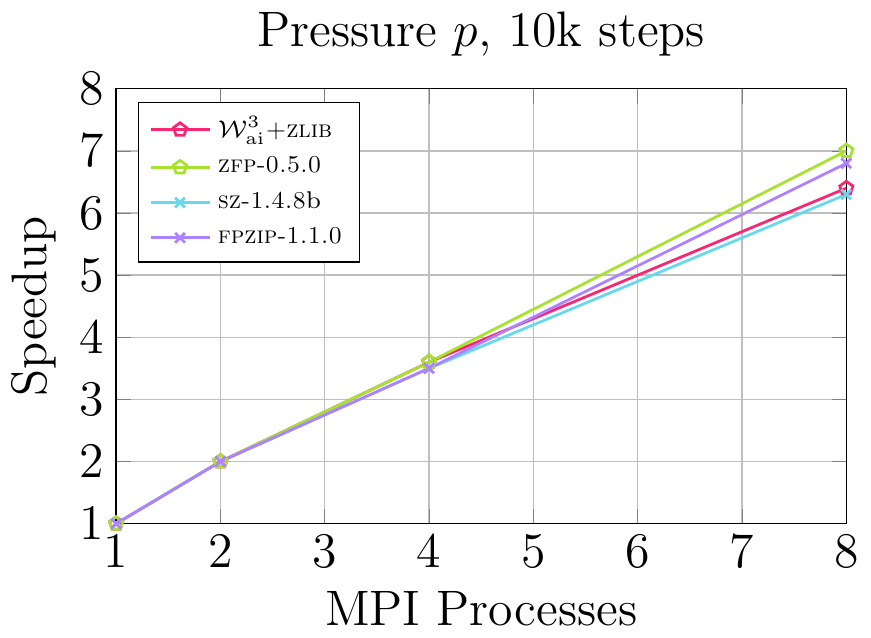}}
    \end{subfigure}

    \begin{subfigure}[t]{0.45\textwidth}
      \centering
      \resizebox{\textwidth}{!}{\includegraphics{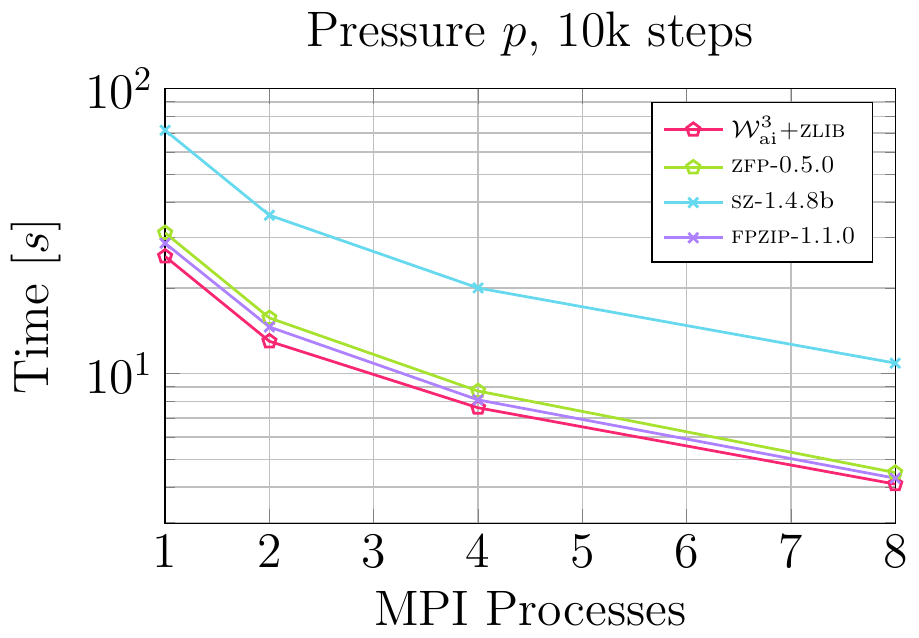}}
    \end{subfigure}%
    \hspace{1em}
    \begin{subfigure}[t]{0.42\textwidth}
      \centering
      \resizebox{\textwidth}{!}{\includegraphics{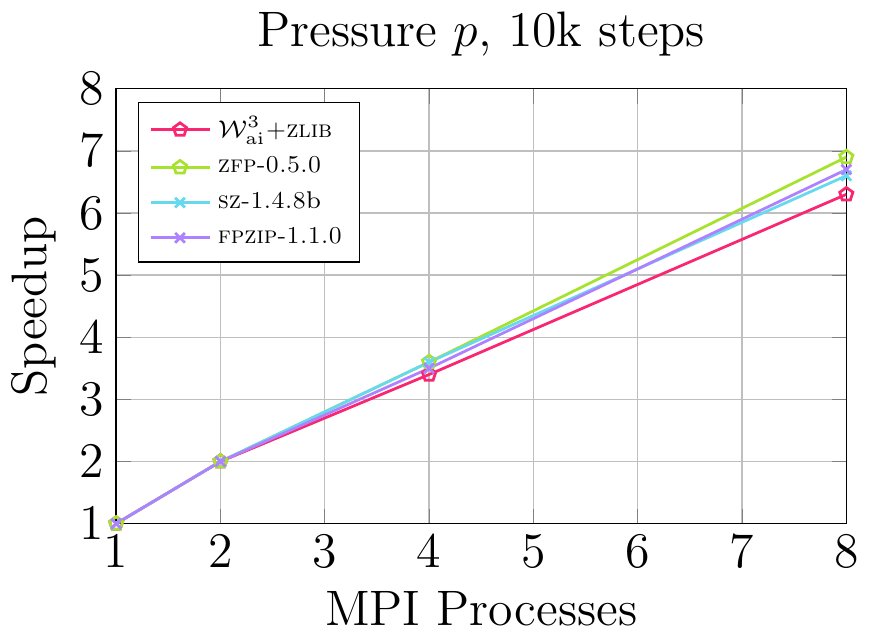}}
    \end{subfigure}

    \caption{Compression time and speedup for four methods on Piz Daint,
    for the same dataset using $512^3$ (top) and $1024^3$ (bottom) grid points.}
    \label{fig:daint2}
  \end{figure}

  Figure~\ref{fig:daint2} depicts the performance of CubismZ using up to 8 MPI processes on a single node of Piz Daint.
  The grid blocks of the dataset are divided evenly among the MPI processes. This configuration allows us to perform measurements for the non-thread-safe floating point compressors integrated into our framework.
  We observe good scaling behavior for all methods and both problem sizes ($512^3$ and $1024^3$).
  The execution times for the smaller problem size are consistent with the compression speeds reported in Table~\ref{tab:speeds}.
  For the bigger problem size, wavelets are significantly faster because the specific error threshold ($\varepsilon = 10^{-3}$) produces fewer coefficients, reducing thus the time spent in the slower ZLIB compression.

  \subsection{Weak scaling}

\begin{figure}[t]
    \centering

    \begin{subfigure}[t]{0.45\textwidth}
      \centering
      \resizebox{\textwidth}{!}{\includegraphics{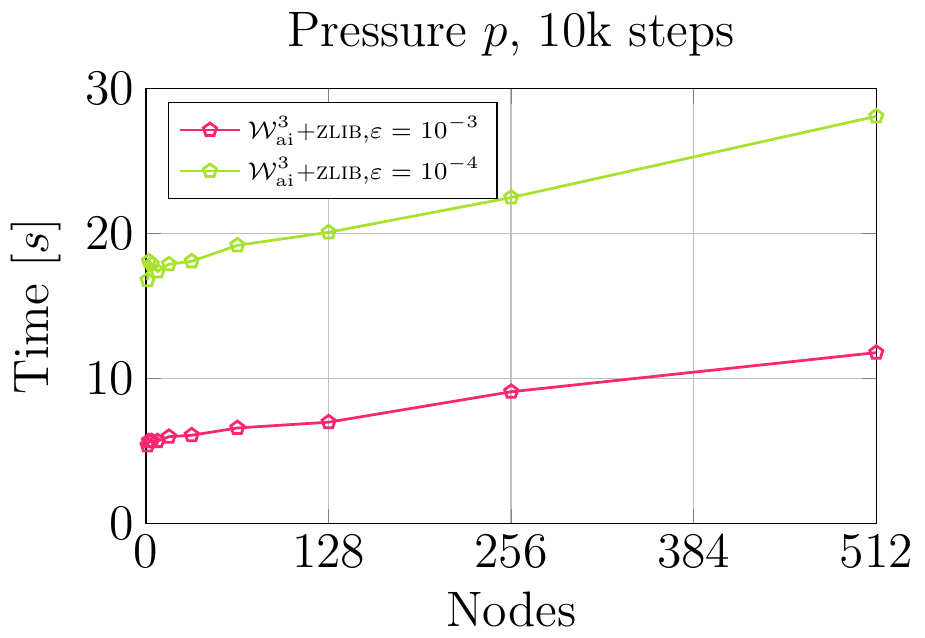}}
    \end{subfigure}%
    \hspace{1em}
    \begin{subfigure}[t]{0.51\textwidth}
      \centering
      \resizebox{\textwidth}{!}{\includegraphics{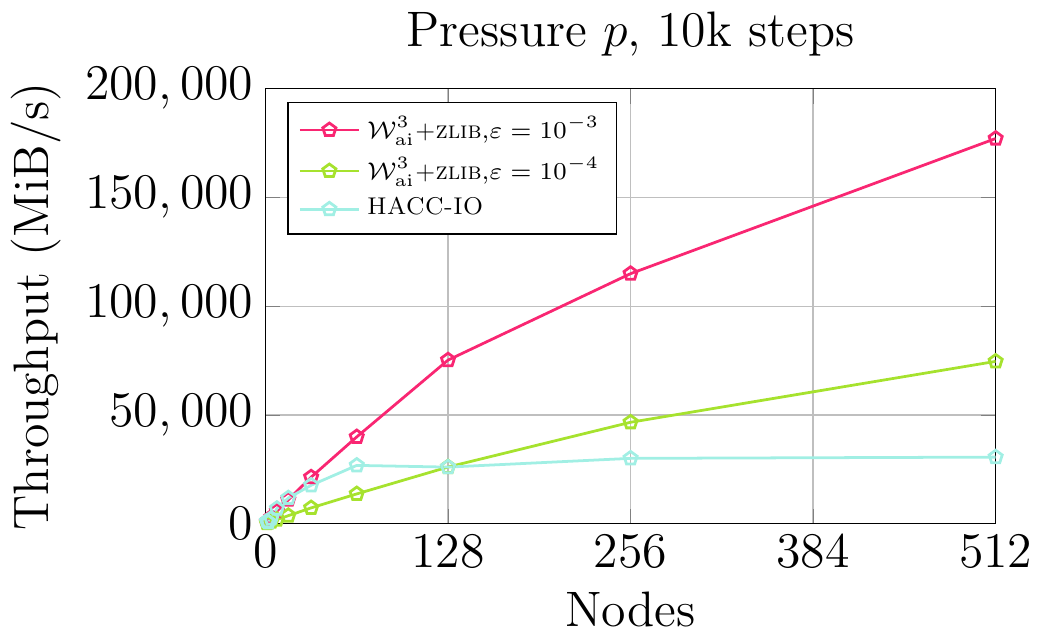}}
    \end{subfigure}

    \caption{Compression time and throughput (including write I/O) for wavelets with ZLIB on Piz Daint.}
    \label{fig:daint3}
  \end{figure}

  We study the weak scaling of CubismZ, for its wavelet-based compression scheme, on up to 512 compute nodes of Piz Daint.
  On each node we utilize a single MPI process with 12 OpenMP threads and compress 4 GB ($1024^3$ points) of pressure $p$. 
  The compression ratio is 57.43 for error threshold $\varepsilon = 10^{-3}$ and 12.01 for $\varepsilon = 10^{-4}$, which result in 71.3 and 341.0 MB of compressed data per node.
  Moreover, the corresponding PSNR values are 93 and 109 dB, respectively.
  Figure~\ref{fig:daint3} depicts the overall time for compressing and writing the data to a shared single file (left) and the equivalent I/O throughput in MB/s (right).
  This experiment represents the real-case scenario where recently updated data are compressed and stored to files as the simulation evolves.
  We observe that the execution time increases with the number of nodes, which is attributed to the communication and I/O overheads induced by the parallel file writing procedure.
  Figure~\ref{fig:daint3}(right) also shows the effective I/O throughput reported by the HACC-IO benchmark, using a single MPI process per node.
  In all experiments, the striping factor of the output file was set to its maximum value (28) for the Sonexion 3000 parallel file system of Piz Daint.
  The nominal peak performance of the file system is 112 GB/s while the effective peak for sequential writes is 81 GB/s,
  measured by the CSCS staff using the IOR synthetic benchmark on 1200 compute nodes, for full machine reservation and empty file
  system~\cite{Alam:2017}.

  \subsection{Large scale simulations}

\begin{figure*}[t]
    \centering
      \resizebox{0.75\textwidth}{!}{\includegraphics{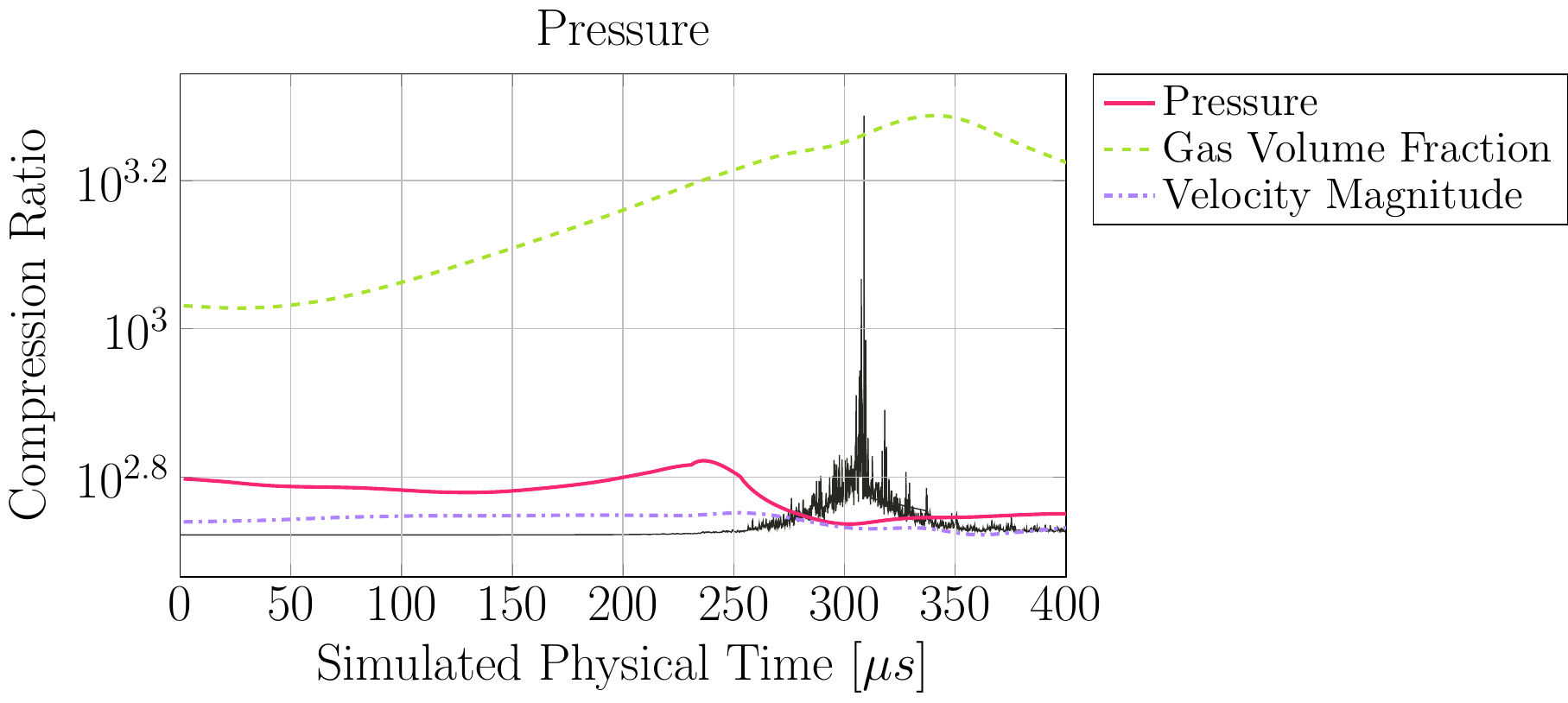}}
    \caption{Compression ratios as a function of time for a large scale simulation with $\mathcal{O}(10^{11})$ grid cells.
    The black solid line shows the local peak pressure and indicates the degree of distortion in the data.}
    \label{fig:mira_compression}
  \end{figure*}

  Figure~\ref{fig:mira_compression} shows compression ratios over time for the production run of a large scale simulation of cloud cavitation involving a spherical bubble cloud with $12500$ cavities.
  The simulation has been executed using $16384$~nodes on the MIRA IBM BG/Q supercomputer at ANL.
  The computational domain has been divided into $7.1$~Million blocks with a block size of $32$~cells, where each node operates on a subset of $432$ blocks.
  This number is smaller than the $4096$ blocks that can almost fully utilize a single node, because we exploit the strong scaling capabilities of our solver to reduce the total execution time.
  The discretization yields a resolution of $\mathcal{O}(10^{11})$ grid cells in the computational domain.

  During the simulation the I/O routine has been executed $262$ times, generating compressed snapshots of three scalar fields of uncompressed size $884$~GB each.
  The compression scheme combines third order averaging interpolating wavelets with ZLIB, while the error threshold has been adjusted for each QoI to yield appropriate quality of data visualizations, i.e., PSNR values in the range of 100-120 dB.
  The achieved compression ratios for the pressure $p$, the gas volume fraction $\alpha_2$ as well as the velocity magnitude $\lVert\bm{U}\rVert$ are depicted in Figure~\ref{fig:mira_compression}.
  The higher ratios compared to those observed in the previous experiments are attributed to the configuration of the simulation, according to which the initial bubble cloud covers a smaller part of the domain.

  Each time, $2652$~GB of data are compressed and written in approximately 14 seconds.
  This corresponds to an equivalent I/O throughput of 190 GB/s on 262'144 cores, which is in accordance with the results reported in~\cite{Bui:2014} for a HDF5-based approach.
  Our approach optimizes I/O performance through data compression but also yields compression ratios of 100x or higher, significantly reducing storage requirements.
  The total overhead due to I/O amounts to only $2\%$ for this particular simulation, with a total size of $1.1$TB for all of the compressed output.
  Moreover, $11$~compressed restart snapshots, containing 7 solution fields, were generated during the simulation.
  Their compression ratio using lossless FPZIP ranged from 2.62x to 4.25x.

\section{Conclusion and outlook}
  \label{sec:conclusions}
  We presented CubismZ, a framework for parallel and distributed ex situ and in situ compression of scientific 3D datasets.
  The software includes a wavelet based compression scheme that yields high compression rate and peak signal-to-noise ratio, and has negligible impact on the execution time of large scale simulations.
  It also incorporates state-of-the-art floating point compressors allowing for a straightforward comparison using datasets generated from various simulation fields.
  The results show that the performance of the data compression schemes for our flow simulations depends on both, the QoI to be compressed and the requested accuracy.
  For most of the QoIs, the wavelet-based data compression yields the highest ratios for PSNR values in the range of 100-120 dB, which is suitable for our visualization needs.

  Our current work includes experimentation with second-generation wavelet types, which can provide better PSNR values and compression ratios.
  Further improvements include adaptive control of the error threshold and extension of the code so that  CubismZ can be easily integrated into third-party simulation codes.

\section*{Acknowledgments}
  This work was supported by the ERC Advanced Grant No 34117 (FMCoBe), the DOE INCITE project CloudPredict, the PRACE projects PRA091 and Pra09 2376, and the Swiss National Supercomputing Center (CSCS) projects s500 and s754\footnote{\url{http://www.gauss-centre.eu/gauss-centre/EN/Projects/CSE/2016/Koumoutsakos_cloud_cavitation_collapse.html}}.
  The authors would like to thank Dr. Diego Rossinelli, Dr. Ursula Rasthofer and Prof. Petros Koumoutsakos for providing feedback during manuscript drafting.
  
\section{References}


\begin{thebibliography}{0}
\bibitem{1}P. Lindstrom, Fixed-rate compressed floating-point arrays, IEEE Trans. Vis. Comput. Graphics 20 (12) (2014) 2674--2683.
\bibitem{2}S. Di, F. Cappello, Fast error-bounded lossy HPC data compression with SZ, in: IEEE Intl. Parallel and Distrib. Processing Symp., 2016, pp. 730--739.
\bibitem{3}P. Lindstrom, M. Isenburg, Fast and efficient compression of floating-point data, IEEE Trans. Vis. Comput. Graphics 12 (5) (2006) 1245--1250.
\bibitem{4} HDF Group, Hierarchical data format, version 5 (2017).
\end{thebibliography}

\begin{thebibliography}{10}
\expandafter\ifx\csname url\endcsname\relax
  \def\url#1{\texttt{#1}}\fi
\expandafter\ifx\csname urlprefix\endcsname\relax\def\urlprefix{URL }\fi
\expandafter\ifx\csname href\endcsname\relax
  \def\href#1#2{#2} \def\path#1{#1}\fi

\bibitem{Habib:2013}
S.~Habib, V.~Morozov, N.~Frontiere, H.~Finkel, A.~Pope, K.~Heitmann, {HACC}:
  Extreme scaling and performance across diverse architectures, in: Intl. Conf.
  on High Perf. Computing, Networking, Storage and Analysis, 2013, pp.
  6:1--6:10.

\bibitem{Sillero:2011}
J.~Sillero, G.~Borrell, J.~Jim{\'e}nez, R.~D. Moser, Hybrid {OpenMP-MPI}
  turbulent boundary layer code over 32{K} cores, in: 18th European MPI Users'
  Group Conference on Recent Advances in the Message Passing Interface,
  EuroMPI'11, 2011, pp. 218--227.

\bibitem{Lee:2013}
M.~Lee, N.~Malaya, R.~D. Moser, Petascale direct numerical simulation of
  turbulent channel flow on up to 786{K} cores, in: Intl. Conf. for High Perf.
  Computing, Networking, Storage and Analysis, 2013, pp. 61:1--61:11.

\bibitem{Rossinelli:2013}
D.~Rossinelli, B.~Hejazialhosseini, P.~Hadjidoukas, C.~Bekas, et~al., 11
  {PFLOP/s} simulations of cloud cavitation collapse, in: Intl. Conf. on High
  Perf. Computing, Networking, Storage and Analysis, 2013, pp. 3:1--3:13.

\bibitem{Rudi:2015}
J.~Rudi, A.~C.~I. Malossi, T.~Isaac, G.~Stadler, et~al., {An extreme-scale
  implicit solver for complex PDEs: Highly heterogeneous flow in earth's
  mantle}, in: Intl. Conf. for High Perf. Computing, Networking, Storage and
  Analysis, ACM, 2015, p.~5.

\bibitem{Yang:2016}
C.~Yang, W.~Xue, H.~Fu, H.~You, et~al., 10m-core scalable fully-implicit solver
  for nonhydrostatic atmospheric dynamics, in: Intl. Conf. for High Perf.
  Computing, Networking, Storage and Analysis, 2016, pp. 6:1--6:12.

\bibitem{Lofstead:2008}
J.~F. Lofstead, S.~Klasky, K.~Schwan, N.~Podhorszki, C.~Jin, {Flexible IO and
  integration for scientific codes through the adaptable IO system (ADIOS)},
  in: 6th Intl. Wrkshp. on Challenges of Large Applications in Distrib.
  Environments, 2008, pp. 15--24.

\bibitem{Li:2003}
J.~Li, W.-k. Liao, A.~Choudhary, R.~Ross, et~al., {Parallel netCDF: a
  high-performance scientific I/O interface}, in: ACM/IEEE Conf. on
  Supercomputing, 2003, pp. 39--39.

\bibitem{Bicer:2014}
T.~Bicer, J.~Yin, G.~Agrawal, Improving {I/O} throughput of scientific
  applications using transparent parallel compression, in: 14th IEEE/ACM Intl.
  Symp. on Cluster, Cloud and Grid Computing, 2014, pp. 1--10.

\bibitem{Schendel:2012}
E.~R. Schendel, S.~V. Pendse, J.~Jenkins, D.~A. Boyuka, II, et~al., {ISOBAR
  hybrid compression-I/O interleaving for large-scale parallel I/O
  optimization}, in: 21st Intl. ACM Symp. on High-Performance Parallel and
  Distrib. Comput., 2012, pp. 61--72.

\bibitem{Burtscher:2009}
M.~Burtscher, P.~Ratanaworabhan, {FPC}: A high-speed compressor for
  double-precision floating-point data, IEEE Trans. Comput. 58~(1) (2009)
  18--31.

\bibitem{Folk:2011}
M.~Folk, G.~Heber, Q.~Koziol, E.~Pourmal, D.~Robinson, An overview of the
  {HDF5} technology suite and its applications, in: EDBT/ICDT 2011 Workshop on
  Array Databases, 2011, pp. 36--47.

\bibitem{Sakai:2013}
R.~Sakai, D.~Sasaki, K.~Nakahashi, Parallel implementation of large-scale {CFD}
  data compression toward aeroacoustic analysis, Comput. Fluids 80 (2013)
  116--127.

\bibitem{Bui:2014}
H.~Bui, H.~Finkel, V.~Vishwanath, S.~Habib, et~al., {Scalable parallel I/O on a
  Blue Gene/Q supercomputer using compression, topology-aware data aggregation,
  and subfiling}, in: 22nd Euromicro Intl. Conf. on Parallel, Distrib., and
  Network-Based Processing, 2014, pp. 107--111.

\bibitem{Gailly:2004}
J.-L. Gailly, M.~Adler, Zlib compression library (2004).

\bibitem{Alted:2010}
F.~Alted, {Why modern CPUs are starving and what can be done about it},
  Computing in Science and Engineering 12~(2) (2010) 68--71.

\bibitem{Lakshminarasimhan:2013}
S.~Lakshminarasimhan, N.~Shah, S.~Ethier, S.-H. Ku, C.~Chang, S.~Klasky,
  R.~Latham, R.~Ross, N.~F. Samatova, {ISABELA for effective in situ
  compression of scientific data}, Concurr. Comput.: Pract. Exper. 25~(4)
  (2013) 524--540.

\bibitem{Lindstrom:2006}
P.~Lindstrom, M.~Isenburg, Fast and efficient compression of floating-point
  data, IEEE Trans. Vis. Comput. Graphics 12~(5) (2006) 1245--1250.

\bibitem{Lindstrom:2014}
P.~Lindstrom, Fixed-rate compressed floating-point arrays, IEEE Trans. Vis.
  Comput. Graphics 20~(12) (2014) 2674--2683.

\bibitem{Di:2016}
S.~Di, F.~Cappello, {Fast error-bounded lossy HPC data compression with SZ},
  in: IEEE Intl. Parallel and Distrib. Processing Symp., 2016, pp. 730--739.

\bibitem{Hadjidoukas:2015a}
P.~E. Hadjidoukas, D.~Rossinelli, B.~Hejazialhosseini, P.~Koumoutsakos, {F}rom
  11 to 14.4 {P}{F}{L}{O}{P}s: {P}erformance {O}ptimization for {F}inite
  {V}olume {F}low {S}olver, in: 3rd Intl. Conf. on Exascale Applications and
  Software (EASC'15), University of Edinburgh, Edinburgh, 2015, pp. 7--12.

\bibitem{Fischer:2008}
P.~F. Fischer, J.~W. Lottes, S.~G. Kerkemeier, nek5000 web page, Web page:
  \url{http://nek5000.mcs.anl.gov}.

\bibitem{Desjardins:2008}
O.~Desjardins, G.~Blanquart, G.~Balarac, H.~Pitsch, High order conservative
  finite difference scheme for variable density low mach number turbulent
  flows, Journal of Computational Physics 227~(15) (2008) 7125--7159.

\bibitem{Donoho:1992}
D.~L. Donoho, Interpolating wavelet transforms, Preprint, Department of
  Statistics, Stanford University 2~(3) (1992) 1--54.

\bibitem{Cohen:1993}
A.~Cohen, I.~Daubechies, P.~Vial, Wavelets on the interval and fast wavelet
  transforms, Appl. Comput. Harmon. Anal. 1~(1) (1993) 54--81.

\bibitem{Masui:2015}
K.~Masui, M.~Amiri, L.~Connor, M.~Deng, et~al., A compression scheme for radio
  data in high performance computing, Astron. Comput. 12 (2015) 181--190.

\bibitem{SPDP:2017}
M.~Burtscher, S.~Claggett, {SPDP compression algorithm v1.0},
  \url{http://cs.txstate.edu/~burtscher/research/SPDPcompressor/} (2017).

\bibitem{LZMA:2017}
{XZ Utils}, \url{https://tukaani.org/xz/} (2017).

\bibitem{LZ4:2017}
{LZ4}, \url{https://lz4.github.io/lz4/} (2017).

\bibitem{ZSTD:2017}
{Zstandard}, \url{http://facebook.github.io/zstd/} (2017).

\bibitem{Iverson:2012}
J.~Iverson, C.~Kamath, G.~Karypis, Fast and effective lossy compression
  algorithms for scientific datasets, in: 18th Intl. Conf. on Parallel
  Processing, Euro-Par'12, Springer-Verlag, Berlin, Heidelberg, 2012, pp.
  843--856.

\bibitem{Alam:2017}
S.~Alam, N.~Bianchi, N.~Cardo, M.~Chesi, et~al., {An operational perspective on
  a hybrid and heterogeneous Cray XC50 system}, in: Cray User Group, 2017, pp.
  1--11.

\end{thebibliography}


\end{document}